# A Taxonomic Review of Adaptive Random Testing: Current Status, Classifications, and Issues

Jinfu Chen[1], Hilary Ackah-Arthur[1], Chengying Mao[2], and Patrick Kwaku Kudjo[1]

[1]School of Computer Science and Communication Engineering, Jiangsu University, Zhenjiang, 202000, China
[2]School of Software and Communication Engineering, Jiangxi University of Finance and Economics, Nanchang, 330013, China

Corresponding author: Jinfu Chen (e-mail: jinfuchen@ujs.edu.cn).

This work is partly supported by National Natural Science Foundation of China (NSFC grant numbers: U1836116, 61462030, and 61762040), the project of Jiangsu provincial Six Talent Peaks (Grant number: XYDXXJS-016), and the Graduate Research Innovation Project of Jiangsu Province (Grant number: KYLX16_0900).

**ABSTRACT** Random testing (RT) is a black-box software testing technique that tests programs by generating random test inputs. It is a widely used technique for software quality assurance, but there has been much debate by practitioners concerning its failure-detection effectiveness. RT is argued to be possibly less effective by some researchers as it does not utilize any information about the program under test. Efforts to mainly improve the failure-detection capability of RT, have led to the proposition of Adaptive Random Testing (ART). ART takes advantage of the location information of previous non-fault-detecting test cases to enhance effectiveness as compared to RT. The approach has gained popularity and has a large number of theoretical studies and methods that employ different notions. In this review, our goal is to provide an overview of existing ART studies. We classify all ART studies and assess existing ART methods for numeric programs with a focus on their motivation, strategy, and findings. The study also discusses several worthy avenues related to ART. The review uses 109 ART papers in several journals, workshops, and conference proceedings. The results of the review show that significant research efforts have been made towards the field of ART, however further empirical studies are still required to make the technique applicable in different test scenarios in order to impact on the industry.

## I. INTRODUCTION

The software industry has two basic problems that require very significant improvement, namely the poor qualities of software and the high development cost [1]. Software testing improves the quality of software [2] and is a very significant activity in any software developments process, as it consumes an approximate of 30% to 50% of a project's budget [3]. There are many software testing techniques [4-10]. Among the many software testing techniques, Random Testing (RT) [4] is a basic and useful black-box [11] testing technique. It is a unit testing technique and plays a crucial role in several testing methods. RT has also been successfully applied in many real-world applications [12-18]. For instance, Forrester and Miller [16] employed the concept RT to test the robustness of Windows NT applications. However, the effectiveness of RT has been questioned by some researchers [19, 20]. There have been controversies as to whether RT is an effective technique to detect software failures.

On the other hand, some researchers have empirically observed that several program faults contribute to failures that form contiguous regions within the input domain [21-24]. These contiguous failure regions showed that other areas of the input domain where the program produces correct outputs (non-failure regions) will also be contiguous [21, 25, 26]. They realized that the presence of contiguous failure regions in the input domain can be more beneficial in improving the RT technique in terms of its failure-detection effectiveness; hence, the proposition of Adaptive Random Testing (ART) [27].

ART is essentially a random testing technique, but with a mechanism which employs the location information of previous test cases in an attempt to widely spread test cases over the input domain. In some empirical studies[1], ART required 50% fewer test cases to detect the first failure when compared with RT. ART technique has also been used in some real-life programs [28-31]. However, the even spread of test cases in ART results in higher computational overhead and affects its efficiency in detecting faults [32]. Over the years, several approaches based on ART concept have been proposed, with the aim of considerably reducing the computational overhead of ART, and improving on its already high fault-detection capability [33-36].

The algorithms of the proposed ART approaches use various notions to attain the goal of even spread. For example, 'exclusion' and 'partitioning' are two varying notions to evenly spread test cases. Some notions to evenly spread test cases have focused on using various features of test cases to select the best candidate. Additionally, various researches have tried to enhance test case selection by combining two or more notions [37-39] or using some generic algorithms [40] to achieve even spread. Although several study contributions to achieve the goal of even spread in ART using varying notions have been made, there is no formal categorization of the study contributions and there is less knowledge of the trends of studies on ART over the years.

In summary, ART has been broadly studied since it was first proposed by Chen *et al.* [41], as much studies have been done seeking to turn the technique into a practical testing approach. However, not much work has been done to highlight the trends of contributions to these ART studies over the years. Apart from a few PhD theses [42, 43], review papers focusing on ART have seldom been published. The objective of this article is to review some of the existing studies on ART. In this review, we contextualize, classify, and assess existing ART studies and proposed methods with a focus on their motivation, strategy, and findings. The study also discusses several worthy avenues related to ART. The results of our review may be used as a reference for further studies on software testing, especially for ART studies. Our review also significantly expands on the knowledge of software engineering.

In the following sections, Section II presents the background of software testing, the ART technique, common patterns of program failure, and performance measures normally used to evaluate ART. Thereafter, the protocol used for the review, comprising of resources and the strategies adopted for the review process, are described and elaborated in Section III. Research questions which draw some comparisons with the various literature and methods of ART are also defined in Section III. This is followed by Section IV, which presents the results obtained from the review of selected studies with respect to defined research questions. The results include categorization of ART study types, the number of ART studies by year, author contributions, ART methods for numeric programs and the number of ART methods by year. Section V discusses the findings by providing answers to the research questions, as well as discusses some open issues and then points out future directions in the study area. Finally, Section VI draws a conclusion to this review.

## II. OVERVIEW OF TESTING

Software testing is a process of executing a program with the intent of finding software bugs or errors in order to satisfy test requirements. Test requirements are specific things that must be satisfied or covered; e.g., statement coverage requires the reaching of statements, mutation requires the killing of mutants, and data flow testing requires the execution of DU pairs. A testing method serves as a guide to a tester in a testing process by incorporating a test criterion. A basic black-box testing technique is random testing.

### A. RANDOM TESTING

Random testing is a testing technique that selects test cases randomly according to either a uniform distribution or based on the operational profile. The motivation is that, if we do not have any information (such as error pattern or intrinsic properties) about the program under test, then each test case is as likely as the others are to be able to detect failures. Hamlet [4] pointed out that, the main advantages of RT include the availability of efficient algorithms to generate its test cases, and its ability to infer reliability and statistical estimates.

RT can be used as a standalone or applied with other testing methods, and due to its conceptual simplicity and efficiency for test case generation, it has been a commonly employed to find failures in various programs [12, 14-16, 18]. However, it is controversial as to whether it qualifies as an effective testing method or not [19, 20]. Myers [19] criticized that RT may be the 'least effective' method as it does not make use of any information about the program under test. Thayer *et al.* [44] however, recommended the use of random testing at the final testing of a program.

In a study by Chan *et al.* [23], it was established that in addition to the failure rate, the performance of a testing strategy also depends on the geometric pattern of the failure-causing inputs. This prompted the investigation to access whether the performance of RT can be improved by considering the patterns of failure-causing inputs. It was evident in [20] that RT does not ensure an even distribution of test inputs over the input domain and may not discover corner cases than other testing approaches. In addition, if the distribution of failing inputs follows some specific patterns, then this information could be exploited to give a higher probability of sampling inputs that are more likely to detect faults. This may result in the development of a modified version of RT that takes advantage of the patterns of failure–causing inputs.

### B. PATTERNS OF FAILURE-CAUSING INPUTS

Many researchers have independently conducted investigations on failure patterns, and have had similar observations that inputs that cause failure often tend to cluster into contiguous regions (*failure regions*). Van der Meulen *et al.* [45] collected a large number of programs written in various languages by different authors and attempted to characterize the faults in these programs. One of their observations is that failure-causing inputs are often clustered into contiguous failure regions. Bishop [21], in an empirical study, found that the large majority of faults detected caused contiguous failure patterns with "sharp" edges—in the terms of Chan *et al.* [23], "strip" or "block" failures. He furthermore argued that some of the edges were likely to be aligned with "contours of equal output value for the function upstream of the error". Ammann and Knight [22] found similar patterns in the small-scale study of a hypothetical missile control programs, that 'at the resolution used in scanning', failure regions of these programs are 'locally continuous'. Finelli [25] performed an extensive experiment with an objective of characterizing software failure processes using different research categories. These categories consistently observed that their programs generated errors each time inputs are chosen from some contiguous regions of the input space. White and Cohen [24] also investigated a common program error, namely domain error (a fault that is located in some predicate of the program under test) and found that domain errors normally result in contiguous failure regions. A very recent study by Sinaga [46] also reports of test inputs with similar coverage residing in contiguous failure regions. Although all of these studies made comparable observations about the behavior of program failure, the motivation for their respective studies quite different.

---

[1] Note: The term studies as used in this review also refers to the number of paper retrieved.

Chan *et al.* [23] observed in a study that certain types of frequently occurring errors in programs were likely to produce failures that form regular "patterns" throughout the input domain, and the patterns of failure-causing inputs (failure patterns) can also affect the performance of some partition testing strategies. They categorized the failure patterns into three categories: the *block*, the *strip*, and the *point* patterns. To illustrate this, we can assume a two-dimensional input domain, as shown in Figure 1 below.

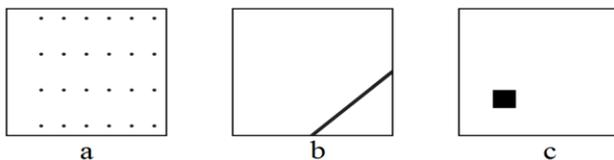

FIGURE 1. Classifications of the patterns of failure-causing inputs: (a) point, (b) strip, and (c) block patterns. The borders represent two-dimensional input domains, and the filled regions denote the failure-causing inputs that are the failure patterns

For the point pattern, the failure-causing inputs have the characteristic of either stand-alone points or form regions of a very small size which are scattered over the whole domain with each region possibly representing only a single test case. For the strip pattern, the failure-causing inputs have the characteristic shape of a narrow elongated strip. A typical example of this failure pattern is White and Cohen's domain errors [24]. The block pattern is mainly characterized by a concentration of the failure-causing inputs in either a single or a few contiguous compact regions of the program's input space. Chan *et al.* [23] noted that these points were sometimes spread in a regular pattern throughout the input domain. They claimed that the block and strip failure patterns were likely to be more common than the point pattern. Example 1, Example 2, and Example 3 show pseudo-code examples of program snippets containing specific errors that lead to the failure patterns, respectively.

---

Example 1: A program fault that results in block failure pattern.
    INTEGER X, Y, Z
    INPUT X, Y
    IF (X > 0 AND X < 10 AND Y > 0 AND Y < 10)
    Z = X  /*correct statement: Z = 2*X */
    ELSE
    Z = 2*Y
    OUTPUT Z

Example 2: A program fault that results in strip failure pattern.
    INTEGER X, Y, Z
    INPUT X; Y
    IF (Y <= 0)  /* correct statement: IF(Y <= 1) */
    Z = X - 2Y
    ELSE
    Z = X + 2Y
    OUTPUT Z

Example 3: A program fault that results in point failure pattern.
    INTEGER X, Y, Z
    INPUT X, Y;
    IF (X mod 4 = 0 AND Y mod 4 = 0)
    Z = X – Y  /* correct statement: Z = X + Y */
    ELSE
    Z = X * Y
    OUTPUT Z

---

### C. ADAPTIVE RANDOM TESTING

On the basis that contiguous failure regions are common, two test cases that are close to each other have a higher probability of exhibiting the same failure behavior as compared to two test cases that are widely spaced. Therefore, given a choice between a point *A* that is close to other points that have already been tested but have not detected a failure, and a point *B* that is further away; the point that is further away is more likely to reveal a fault. This is the notion that guides Adaptive Random Testing in its test case selection.

ART is a random-based test data generation and selection technique that enhances the effectiveness of tests over pure random testing (RT). The concept of ART which was first introduced by Chen *et al.* [41] based on a fault-based random testing strategy proposed by Mak [47], and has been designed to detect common failure patterns better than pure RT. ART refers to those approaches to software testing which are based on RT but include some additional mechanism to encourage a more widespread and even distribution of test cases over the input domain. The intuition of ART is to spread random test cases evenly over the input domain to increase the likelihood of finding failures (especially for non-point types of failures) with lesser number of test cases than ordinary RT. To generate a new test case, it is necessary to ensure that the new test case is farther away from all cases that have been generated previously. One way to achieve this is to generate a number of random test cases and then choose the "best" one among them. That is, to try to distribute the selected test cases as spaced out as possible.

---
**Algorithm 1** *Basic ART (FSCS-ART)*
1: Set $n = 0$.  //$n$ represents the number of test cases generated so far.
2: Set $\mathbf{E} = \{\}$.  //To store executed test cases.
3: Randomly select a test case $t$ from the input domain (according to uniform distribution).
4: Increment $n$ by 1.
5: If $t$ reveals a failure, go to Step 9; otherwise, store $t$ in $\mathbf{E}$.
6: Randomly generate $k$ tests to construct $\mathbf{C}$ (according to uniform distribution).
7: For each $c_i \in \mathbf{C}$, calculate the distance $d_i$ between $c_i$ and its nearest neighbor in $\mathbf{E}$.
8: Find $c_b \in \mathbf{C}$ such that its $d_b \geq d_i$ where $n \geq i \geq 1$.
9: Let $t = c_b$ and go to Step 3.
10: Return $n$ and $t$, and EXIT.

---

For instance, the basic algorithm of ART, the *Fixed-size-candidate-set* (FSCS) [41], [27] makes use of two sets of disjoint test cases—the *candidate set* (C) and the *executed set* (E). The candidate set is a fixed set of test cases that are randomly selected from the input domain without replacing them; whereas the executed set refers to the set of previously executed test cases that failed to reveal any failure. Algorithm 1 provides a formal description of the basic ART.

The algorithm begins with an empty executed set and selects an initial test case randomly from the input domain of the software under test. If it does not reveal any failure, an initial executed set is formed by this test case as the only element. Rather than randomly generating a test case from the input domain each time as RT does, the algorithm randomly generates a fixed number of test cases to form the candidate set. It then chooses the farthest candidate element to all the elements in the executed set (i.e. candidate test that is farthest away from the already used inputs) as the next test input. This input is then executed using the software under test; if it does not reveal a failure, the test input is put into the executed set. The remaining elements of the candidate set are discarded once a test case has been chosen and a new candidate set is constructed. The process of incrementing the executed set with a selected element from each candidate sets is repeated until a failure is revealed by a failure-causing input, or until the stopping condition is reached. Figure 2 illustrates the basic process of ART.

When computing the farthest candidate element in Figure 2, the *Euclidean distance metric* is normally used to calculate distances between two numerical test cases. That is, for an $n$-dimensional input domain, the distance between two test cases $a$ and $b$ whose inputs are $a_i$ and $b_i$ respectively, for $i \in \{1, ..., n\}$, is:

$$dist(\mathbf{a}, \mathbf{b}) = \sqrt{\sum_{i}^{n}(a_i - b_i)^2} \qquad (1)$$

Chen *et al.* [27] performed experiments using 12 error-seeded published programs, all involving numerical computations. Their results were very encouraging showing that ART does outperform ordinary RT significantly (for even as much as 50%) for the set of programs studied. This provided evidence of its improved effectiveness over RT.

The ART technique has mainly three types of notions to achieve even spread of test cases: *Distance* notion, *Exclusion* notion, and *Partition* notion. The ART approaches that employ the Distance notion compute distances to potential test candidates within the boundary of the input domain, in their test case selection strategies. These approaches mostly provide variants of the basic ART algorithm mainly to further improve failure detection effectiveness. Approaches that employ the Exclusion notion restrict the selection of test cases to regions further away from previously executed test cases to achieve even spread. The first variant of these approaches [33] had comparable failure detection effectiveness with similar time complexity $O(n^2)$ as the basic ART. Although the Exclusion notion utilizes distance computation in determining qualified test cases, its classification is based on the definition of restricted regions. Partition approaches divide the input domain and samples test cases from the different partitions. Most variations of Partition approaches [48], [49] have the advantage of reducing computational cost than the basic ART. Several ART approaches that employ the Partition notion combine other strategies in the generating test cases from the different partitions [39], [35, 36]. Besides that, researchers have developed other diverse ART approaches with different notions from the three main ART notions but obey the principles of ART to achieve randomness and even spread in test case generation. In this study, we henceforth refer to the notion of these other diverse test generation approaches as the *Alternative* notion, for convenience.

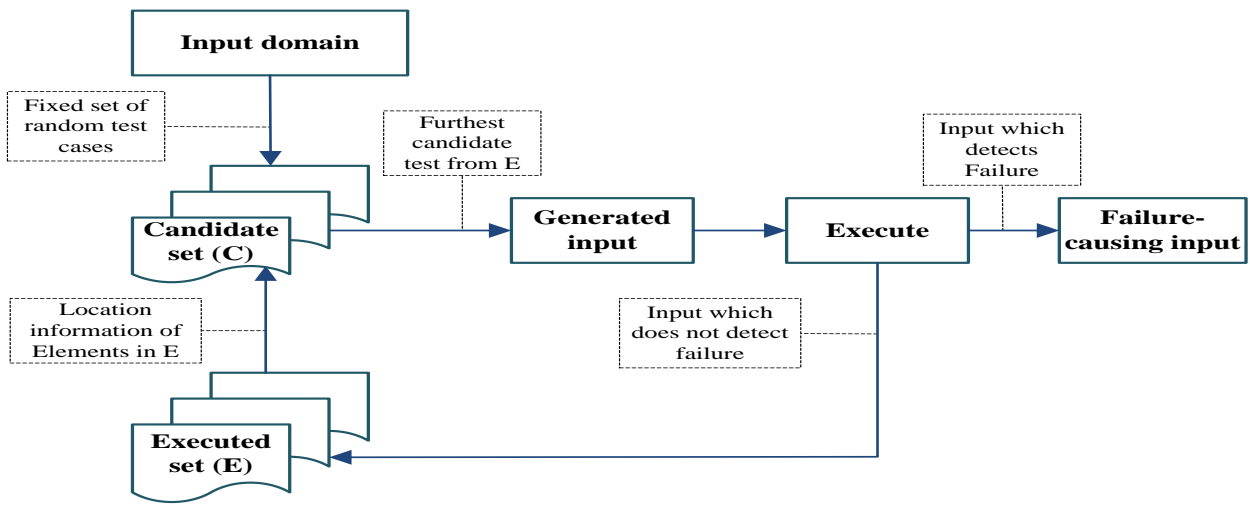

FIGURE 2. **Basic Process of ART.**

Other attempts to utilize the continuity of failure regions, by employing various techniques to attain the spread of test cases evenly, include *Anti-random Testing* [50] and *Quasi-random Testing* [42].

Quasi-Random Testing also uses a class of sequences, known as quasi-random sequences, that intrinsically filters or restricts randomly selected test cases to conform to the "separateness" requirements. It differs from ART as the use of sequences does not ensure randomness of test case selection. Chen and Merkel [51] discuss that the quasi-random sequences could be randomized based on two rotating methods (Cranley-Patterson Rotation and Owen's Scrambling Method), but does not bring much randomness into the sequence and incremental generation of test cases is a problem. A further improvement in the randomness by Liu and Chen [52] has resulted in better randomness to test cases and incremental generation of test cases similar to ART methods.

Anti-random testing is not an ART technique as there are a lot of major differences between them. Anti-random testing is more or less exclusively deterministic, as the selection of the first test case presents the only non-determinism in the set. Additionally, the number of required test cases must be chosen beforehand, which is different from the flexible incremental test case generation provided by ART.

The different techniques perform best under varying circumstances. For instance, some techniques provide lower selection costs or an increase in performance when applied to high dimensional input domains.

### D. EVALUATION MEASURES

Random testing methods are normally implemented either with or without replacement of executed test cases [24]. Analytical studies of RT largely assume random test case selection with replacement [8, 20, 24, 53-55], due to its readily available mathematical model and hence its easy analysis. Also, replacement of test cases is very common for testing strategies where the cost of executing test cases is lower than the cost of checking for duplication (as is usually the case for RT). Practitioners have always criticized the assumption of test case selection with replacement, since repeating test cases is not naturally the best practice. In addition, the selection of test cases randomly without replacement very much reflects reality. Most ART studies assume that a random selection of test cases follow a uniform distribution and without replacement. The notation employed by Chen and Yu [56], refers to elements of an input domain that do not produce correct outputs as *failure-causing inputs*. For an input domain $D$, let $d$ represent the size of the $D$, $m$ represent the number of failure-causing inputs, and $n$ represent the number of test cases, respectively. Therefore, we define the sampling rate ($\sigma$) and failure rate ($\theta$) as $n/d$ and $m/d$, respectively.

Two effectiveness measures that have been used in earlier studies of RT are the *P-measure* which is the probability of detecting at least one failure, and the *E-measure* which is the expected number of failures detected. With RT, the E-measure for conducting $n$ tests is $n\theta$, and that of P-measure is $1-(1-\theta)^n$. One relationship between these measures is that when the failure rate and $n$ are sufficiently small, the P-measure and E-measure closely approximate each other [57]. Although these two measures have been accepted by many, they have come under some criticisms. One main issue of using P-measure is the lack of a distinction made when detecting different numbers of failures [27, 43]. The problem with E-measure is that a higher E-measure value does not essentially mean more faults are found or more distinct failures are detected. P-measure and E-measure have been widely used in the literature in the light of their criticisms.

In addition to these two evaluation measures, Chen *et al.* [27] proposed a new measure referred to as *F-measure*, which is the expected number of test cases required to detect the first failure. The F-measure for random selection of test cases with replacement is equal to $1/\theta$, or equivalently $d/m$. The effectiveness of a testing strategy is more viscerally reflected by the F-measure since the measure is more intuitively appealing, and a better match for testing practice when the discovery of a bug causes the testing to be suspended whilst the fault is located [47]. F-measure is more appropriate for evaluating the effectiveness of ART methods. Factors to consider when evaluating the effectiveness (F-measure) of an ART method are the target area (in Restriction) [34], failure pattern or region (in all ART methods), dimension of input domain [58], failure rate [36], and number of test cases [59]. Chen *et al.* [60] conducted an extensive simulation to study the F-measure of ART using various situations. It was found that many ART algorithms obtain smaller F-measures either when the input domain is low in dimension, when the failure region is very compact, or when the number of failure regions is smaller. Practically, testing is normally stopped when a failure is detected and resumed only after the detected fault is fixed. Hence, the F-measure is more intuitive from a practical point of view, and the reason for its adoption as the main effectiveness measure for ART. In general, given a program under test, a testing technique is very effective in detecting failures if it exhibits higher P-measure, higher E-measure or lower F-measure. Empirical studies show that there is a trade-off between effectiveness and computational complexity of any testing approach [61]. Therefore, in a bid to improve the effectiveness of detecting a failure for ART methods, researchers must devise testing strategies that do not sacrifice computation time, and vice-versa.

## III. REVIEW METHODOLOGY

The current investigation was undertaken using procedures outlined in Kitchenham's guidelines [62] and involved an electronic search of multiple online databases. The goal of this review is to analyze scientific papers related to Adaptive Random Testing, focusing on the trends of research contributions and proposed methods. We developed a research protocol to guide and ensure quality in the entire review process. The details of the research protocol we used to search and select related papers for this review are shown in Figure 3.

### A. RESEARCH QUESTIONS

In order to effectively review the selected ART studies; the following research questions (RQs) are addressed by this study:

- RQ1**:** *What are the trends of contributions in ART ART researches to date?*
- RQ2: *What are the proposed variations of ART methods for numeric programs and their characteristics?*
- RQ3: *What are the trends in contributions and development in ART methods for numeric programs?*

We designed the first research question (**RQ1**) to determine the research trends and contributions to ART over the years. Here, the contributions refer to the number of research articles in the area of study. This can enable us to obtain the general and yearly research contributions of ART to date and can provide information on author contributions to the field. We designed the second research question (**RQ2**) to enable us to classify the various ART methods based on the notions they employ to achieve even spread of test cases and to evaluate the characteristics and commonalities of the proposed methods in the respective classifications. The third research question (**RQ3**) was designed to evaluate the contributions to the varying ART methods according to their classification and the yearly ART method proposal to date.

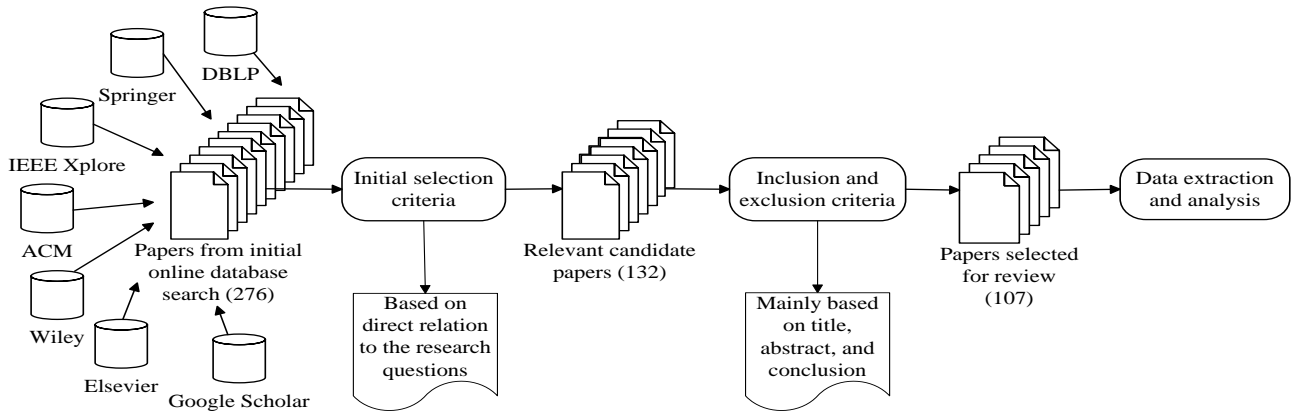

FIGURE 3. **Search and Study Selection Process.**

### B. SEARCH STRATEGY

Our search for related publications focused on the online repositories of the key technical publishers, including Google Scholar, Elsevier online Library, Wiley Interscience, ACM digital library, IEEE Xplore digital library, Springer online library, and DBLP computer science bibliography. We used the keyword search and MeSH terms to identify the related literature from January 2001 to December 2018, concept of ART was first introduced in 2001[41]. For the search terms, the keywords used included "random testing", "adaptive random testing", "software testing", and "testing methods". We extracted the search terms from the RQ. Relevant texts on ART and software testing were searched for inclusion. We included only full papers and letters in the search and filtered out papers not written in English. We also searched for PhD theses that have made a significant contribution to the development of ART. We then performed a backward snowballing [63] by searching the list of references within the papers included for additional relevant but missing papers using the similar keyword rules. We also performed a forward snowballing [63] by searching for papers that cite the current set of papers to reduce the likelihood of missing some relevant papers. By this approach, we performed a "transitive closure" in the literature. We finally obtained 276 potential candidate papers after the electronic search.

*C. SELECTION CRITERIA*

After identifying potential papers, using the search terms described above, we identified those papers that provide direct evidence about the research questions. We defined selection criteria to select more relevant papers based on the research questions. The selection criteria included studies that address topics related to ART such as meta-analysis, methods, and reviews. The use of the selection criteria narrowed the candidate papers to 132 publications within the scope of our review.

*D. SELECTION PROCESS*

Based on the inclusion and exclusion criteria specified, we selected the relevant papers for the review. We assessed every article using their titles, abstracts, conclusions sections and sometimes checking the content of the papers when unsure, in order to judge its relevance. We included articles on ART published between the period of January 1st, 2001 and December 31st, 2018. The inclusion criteria were based on whether the article provides information for addressing the proposed RQs and whether the article is relevant in the ART domain. Also, we considered only articles written in English and included only one version of an article with multiple publications of the same data. Although *Adaptive Random Sequence* (ARS) [64], [65] was originated from the concept of ART [66, 67], the approach differs from ART as the use of sequences does not ensure randomness of test case selection; thus we did not include ARS methods as completely within the scope of ART approaches.

Relevant studies were selected by two researchers and selected and rejected studies were further checked by another researcher. Then, relevant full papers that meet the criteria for inclusion were selected and complete copies obtained for this study. After further assessment of the titles, abstracts, conclusions and sometimes checking the content of the papers when unsure, 109 relevant full papers that meet the criteria for inclusion were selected and complete copies obtained for this study. Summaries of the publication types and the number of studies selected for the review–also referred to as *primary studies* [62]–are illustrated in Table I below.

TABLE I
DISTRIBUTION OF RESEARCH PUBLICATIONS

| Research publications | Number of studies | Percentage |
|---|---|---|
| Journals | 40 | 36.7% |
| Conferences | 62 | 56.9% |
| Workshops | 7 | 6.4% |
| **Total:** | **109** | **100%** |

*E. DATA EXTRACTION STRATEGY*

The 109 papers were all carried through to the data extraction stage. In the data extraction stage, the selected papers were subjected to careful, thorough, and total reading; while related literature was compiled and sorted using *card sorting technique* [68], and relevant data were considered for inclusion in this study.

To effectively differentiate among the selected papers and compile related literature for this review, there was a need to categorize the selected papers based on what notion they propose and achieve. We carefully evaluated the contents of all the papers, in addition to multiple deliberations by the authors, especially with the choice of name for their classification. During the evaluation of the papers, we realized that several studies propose ART methods; therefore we categorized them based on whether they propose ART methods or not. We later realized that even among the studies that propose ART methods, they vary with respect to the notions they employ to randomly and evenly spread test cases. Therefore, we further categorized all the studies that propose ART methods based on the notions they employ to achieve even spread defined in Section II-C. During the evaluation process, new categories emerged and some studies had to be classified again. This procedure was repeated until the categories remained stable. The categorization was based on the idea of "concept matrix" from Webster and Watson [63] for this review. A concept matrix is a logical approach that defines several ''concepts'' (that may be variables, theories, topics, methods, and so on), where all papers are categorized in and therefore represents a classification scheme. We categorized studies that do not propose ART methods as one category, the *Analysis approach*. Studies in this category consist of some theoretical issues on ART such as meta-studies about ART, hypotheses supporting ART, evaluation of features of ART, criticisms of ART, distance metric approaches, thesis, and reviews. Note that all studies that do not introduce any new ART approach but propose some distance metrics and apply them to existing ART approaches [28, 69-74] are included in this category. We then categorized studies that proposed ART methods into four different categories, according to their notions: *Distance*, *Exclusion*, *Partition*, and *Alternative* approaches, as discussed in Section II-C above.

The following relevant data were extracted from each paper and considered for inclusion in this study:

- The source of the paper
- Classification of the study: Type (Distance, Exclusion, Partition, Alternative, Analysis); Scope (numeric)
- Year of publication
- Author information
- Summary of the study including the notion, motivation, and description
- Subject programs used and their descriptions
- Proposed ART methods from each study classification and their algorithms

We utilized a data extraction form for each primary study in order to facilitate the extraction process. All selected studies were shared among all the authors to review and extract data from them. The extracted data were then checked and discussed among the authors to ensure inter-researcher consistency in the data extraction process. The use of one extractor and multiple checkers provides an efficient and useful procedure for data extraction, especially for review of a large number of studies. Although the method of extraction is not consistent with the data extraction guidelines summarized in Kitchenham [62], we found the procedure to be very effective for this study. The procedure of Kitchenham performs the data extraction independently by two or more researchers and the resultant data obtained are compares by the researchers. We shared the selected papers based on our availability in terms of time. We discussed any disagreements until a final data value is agreed upon and recorded. The data extraction and checking processes were repeated for some studies by other researchers when there are disagreements, in order to eliminate misunderstanding and ensure correctness of extracted data. The discussion of any disagreements led to a number of minor changes in the extraction process. We used the Excel spreadsheets to maintain the extracted data, and a library application EndNote X7, to compile bibliographic resources and to effectively retrieve references of related articles. We are reasonably confident of the accuracy of data we extracted from relevant research publications we surveyed.

## IV. RESULTS

The initial electronic search using search strings described above covered the period from January 2001 to December 2018 resulting in 276 titles. After further careful assessment based on both the selection criteria and the criteria for inclusion, we finally identified 109 relevant full papers for this review. Information about the publication types and the number of primary studies selected for this review is presented in Table I above.

This section presents and describes the results of the trend analysis of ART researches, methods, and distance metrics in response to the research questions stated in Section III.

### A. RQ1: TRENDS AND CONTRIBUTIONS OF RESEARCH IN ART

We evaluated the identified papers to answer the research question (RQ1). In order to take a closer look at the trends of research work on ART, we compared the various categories of ART studies defined above in the *data extraction strategy* section (see Section III-E) of this review. The categories of studies we identified are as follows: *Distance, Exclusion, Partition, Alternative, and Analysis*. The first four categories of papers commonly conduct empirical studies (qualitative, quantitative, and mixed) to develop, apply, or validate proposed ART methods. The *Analysis* is the only category which represents studies that do not propose any new ART methods. The related papers we identified under the *Analysis* category are available in [7, 28, 31, 32, 41-43, 59, 60, 66, 69-111].

The goal of the study categories is to allow the authors to analyze the temporal relationship between the developments of research effort towards the various categories by the community. Below in Figure 4, is an illustration of the percentage number of studies we obtained for each category, out of the 109 studies we considered from the beginning of the year 2001 to the end of 2018, inclusive.

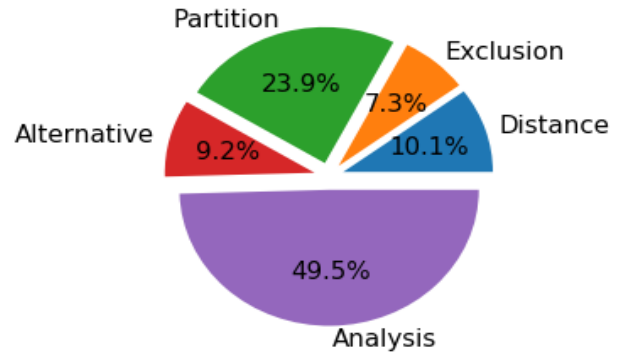

FIGURE 4. **Percentage of ART Studies per Research Notion (2001 - 2018).**

The result in Figure 4 shows that the number of studies in the *Analysis* category is almost half (49.5%) of the total number of select studies considered for this review. For the studies that propose ART methods, the *Partition* category has gained much research (23.9%) more than the *Distance, Exclusion,* and *Alternative* categories. This is followed by the *Distance* (10.1%) and *Alternative* (9.2%) categories; whiles the *Exclusion* category has received the least number of studies (7.3%).

Since the year 2001, studies on ART have focused on finding the optimum improvement over RT; and have introduced several studies that proposed varying methods or analyzed certain properties that can affect results of the ART technique for best performance. To empirically provide an overview of the trend of studies on ART, we calculated the total number of studies for each year. In Figure 5 below, the results of the yearly trend in ART studies are illustrated based on the analysis of the literature collected for this study.

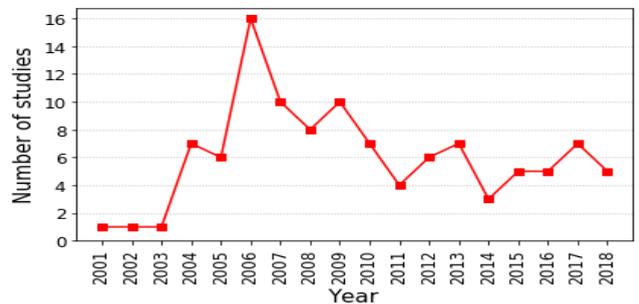

FIGURE 5. **Yearly Trend in ART Studies.**

In a bid to capture up to date ART studies for our investigation, we have included studies published until the end of the year 2018 in this review. The results in Figure 5 show a generally low number of studies in the beginning with a maximum of one paper yearly until the year 2003. The number of yearly studies experienced some increases after the year 2003 with the highest number of ART studies recorded in the year 2006. The yearly number of ART studies has remained constant in the continuous years with an average of five studies per year.

To take a closer look at the research contribution to each category of ART studies we identified, we compared the ART study contributions to each category over the years. The results of the study contributions to each category from the years 2001 to 2018 are illustrated in Figure 6.

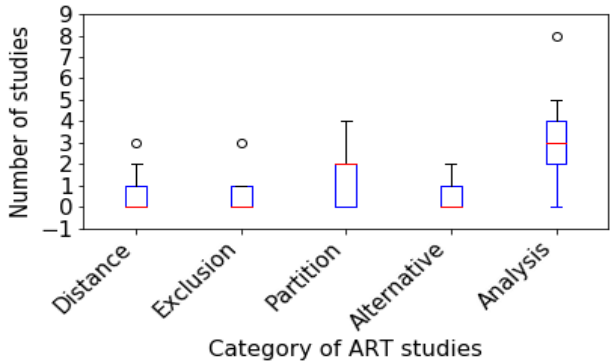

**FIGURE 6. Distribution of ART Studies from 2001 to 2018.**

The results in Figure 6 indicate that there have been consistent studies on the theory of ART. It is also clear that studies in the *Analysis* and *Partition* categories experienced more yearly contributions, although the yearly contributions of the *Analysis* category were consistently high. The contributions to studies in both the *Distance* and *Alternative* categories have also been low but quite consistent over the years, with one study recorded in most of the years. On the other hand, the contributions to studies in *Exclusion* category have rather gained less contribution although it seemed promising with an outlier of three studies in the year 2006. In most of the years, the *Distance, Exclusion* and *Alternative* categories never obtained any study contributions.

The researcher contributions to ART can be an indicator to determine the research impacts and interests in the area of study. We especially made efforts to identify the main researchers in the field of ART and compute the contributions. We identified and populated the names of all authors and co-authors to the literature selected in this study and computed the frequency of appearance of their names in all selected papers. From the literature gathered, we identified 125 authors who have contributed to the research area, with some of them co-authoring in a number of papers. Figure 7 below presents the results of the analysis of the author contribution to ART. Since the illustration of the number of study contributions for all authors identified in the selected literature could increase the complexity of the results, we provide a general summary illustration of the study contributions of all authors in Figure 7(a). We then provide study contributions for the authors, with a focus on only authors or co-authors with contributions in more than two papers, in Figure 7(b).

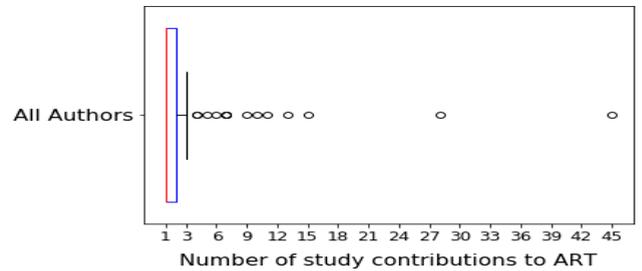

(a)

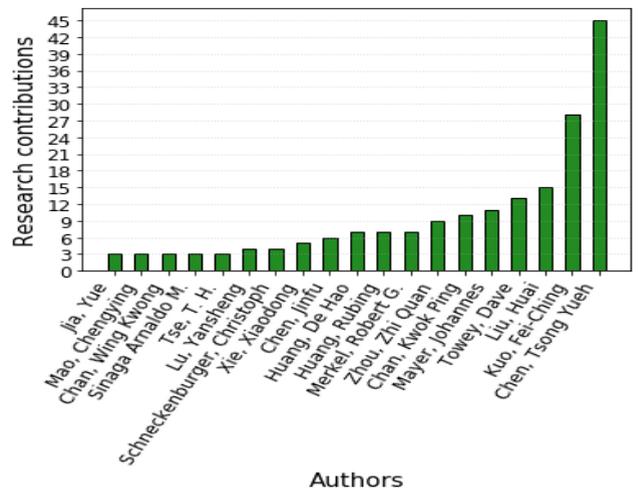

(b)

**FIGURE 7. Author Contributions to ART Studies. (a) general distributions for all authors. (b) distributions for Authors with more than two contributions.**

As shown in Figure 7(a), the majority of the authors have averagely authored or co-authored in one paper. It also shows that very few authors have authored or co-authored in higher numbers of papers. The results in Figure 7(b) show that only 19 authors have contributions in more than two papers in the area of study. Tsong Yueh Chen is the author with the highest number of study contributions in the area of study with 45 co-authored papers. There may be other contributions by the authors to the software testing field, but these results are based on the scope of our study which is ART.

An empirical evaluation of a testing approach using real programs is very essential in determining the quality of the testing approach. Since RT technique works well for simple programs [6], researchers normally perform empirical studies to determine whether ART improves on the effectiveness of RT by using some simple real programs. A number of real programs have been used in the empirical evaluation of ART, though not many. We identified empirical studies that have used real programs in the ART literature. In most empirical studies, all the subject programs (SUT) are seeded with faults belonging to different types of common mutant operations [112]. The fault-seeded programs that are used normally have varying dimensions with some having varying program input lengths.

To provide an overview of the trend on the usage of subject programs for empirical studies on ART, we have collected all the subject programs used in each empirical experiment work from the literature. Table II shows the name, number of lines of code (LOC), program source, description, and the total number of ART research papers that report results for the respective subject programs. Undoubtedly, the definition of LOC in Table II can be problematic; however, the figure is merely intended to be used as a rough indicator. Some of the table entries for LOC and sources of the subject programs are shown as "*Unreported*". This indicates the unavailability of the information among the selected literature. The table is also sorted in descending order, by the number of papers that utilize the respective subject programs in their empirical studies.

TABLE II
SUBJECT PROGRAMS USED IN ART EMPIRICAL STUDIES

| Program Name | *LOC* | Source | Description | Uses |
| --- | --- | --- | --- | --- |
| Bessj | 30 - 200 | Num. Recipes [113] | Computes the ordinary bessel function | 19 |
| Plgndr | 30 - 200 | Num. Recipes [113] | Computes the associated Legendre polynomial equation | 16 |
| Airy | 30 - 200 | Num. Recipes [113] | Computes the airy function | 15 |
| Gammq | 30 - 200 | Num. Recipes [113] | Computes an incomplete gamma function | 15 |
| Bessj0 | 30 - 200 | Num. Recipes [113] | Computes the bessel function for any real value | 14 |
| Erfcc | 30 - 200 | Num. Recipes [113] | Computes a complementary error function | 13 |
| Sncndn | 30 - 200 | Num. Recipes [113] | Computes the Jacobian elliptic functions | 11 |
| Probks | 30 - 200 | ACM (CALGO) [114] | Numeric function | 11 |
| Cel | 30 - 200 | ACM (CALGO) [114] | Numeric function | 11 |
| El2 | 30 - 200 | ACM (CALGO) [114] | Numeric function | 11 |
| Tanh | 30 - 200 | ACM (CALGO) [114] | Numeric function | 9 |
| Golden | 30 - 200 | ACM (CALGO) [114] | Numeric function | 9 |
| Tcas | 141 | Siemens [115] | On-board aircraft traffic collision avoidance system | 7 |
| Replace | 512 | Siemens [115] | Performs pattern matching and replacement | 5 |
| Grep | 9089 | SIR [115] | Unix text-search utility | 4 |
| Triangle | 26 | SIR [115] | Classification program for isosceles and equilateral triangles | 3 |
| Flex | 10124 | SIR [115] | Unix lexical analyser utility | 3 |
| Printtokens | 472 | Siemens [115] | Lexical analyzer | 3 |
| Printtokens2 | 399 | Siemens [115] | Lexical analyzer | 3 |
| Quadratic | *Unreported* | GNU Scientific [116] | Calculates complex roots of quadratic equations | 2 |
| Cubic | *Unreported* | GNU Scientific [116] | Calculate complex roots of cubic equations | 2 |
| Junit | 1500 | JUnit 5 [117] | A unit testing framework | 2 |
| Space | 6199 | SIR [115] | An interpreter for an array definition language | 2 |
| Gzip | 5159 | SIR [115]) | Unix compression utility | 2 |
| Sed | 9289 | SIR [115] | Unix stream text editor utility | 2 |
| Schedule | 292 | Siemens [115] | Priority scheduler | 2 |
| Schedule2 | 301 | Siemens [115] | Priority scheduler | 2 |
| Totinfo | 440 | Siemens [115] | Information measure | 2 |
| Cal | 163 | Unix manual [118] | Prints a calendar for a specified year or month | 2 |
| Comm | 144 | Unix manual [118] | Selects or reject lines common to two sorted files | 2 |

*Table II (Continued)*

| | | | | |
|---|---|---|---|---|
| Look | 135 | Unix manual [118] | Finds words in the system dictionary or lines in a sorted list | 2 |
| Sort | 842 | Unix manual [118] | Sorts and merges files | 2 |
| Spline | 289 | Unix manual [118] | Interpolates smooth curve based on given data | 2 |
| Uniq | 125 | Unix manual [118] | Reports or removes adjacent duplicate lines | 2 |
| Count | 42 | Chris Lott's website [119] | Counts the number of lines, words, and characters in the named files. | 1 |
| Series | 288 | Chris Lott's website [119] | Prints the real numbers from start to end, one per line. | 1 |
| Tokens | 192 | Chris Lott's website [119] | Counts all alphanumeric tokens in inputs and prints their counts | 1 |
| Ntree | 307 | Chris Lott's website [119] | Implements the data structure of an n-ary tree. | 1 |
| Nametbl | 329 | Chris Lott's website [119] | Implements an abstract data structure of a symbol table, as well as its operations. | 1 |
| Select | *Unreported* | Num. Recipes [113] | Finds the smallest element from an array of real numbers | 1 |
| Sin | 99 | [120] | Computes the sine function | 1 |
| Make | 27,879 | SIR [115] | build utility | 1 |
| NANOXML | 7,646 | SIR [115] | XML parser | 1 |
| Trisqure | 168 | SIR [115] | Calculates the type and square of a triangle constructed. | 1 |
| Triangle2 | 41 | SIR [115] | Returns the type of triangle | 1 |
| Median | 20 | Num. Recipes [113] | Basic mathematical routines | 1 |
| Remainder | 48 | [121] | Checks if the remainder is zero or non-zero after a division | 1 |
| Variance | 22 | Num. Recipes [113] | Basic mathematical routine | 1 |
| BubbleSort | 14 | [121] | Basic mathematical routine | 1 |
| Encoder | 65 | Num. Recipes [113] | Basic mathematical routine | 1 |
| Expint | 86 | Num. Recipes [113] | Basic mathematical routine | 1 |
| Fisher | 71 | Num. Recipes [113] | Basic mathematical routine | 1 |
| Siena | 6035 | SIR [115] | Wide-area event notification system | 1 |
| DRUPAL | 336,025 | [122] | Web framework | 1 |
| BUSYBOX | 189,722 | [123] | UNIX utilities | 1 |
| LINUX | 12,594,584 | [123] | Operation system | 1 |
| GCD | 55 | Unreported | Calculates the greatest common divisor | 1 |
| LCM | *Unreported* | Unreported | Calculates the least common multiplier | 1 |
| Checkeq | 90 | Unix manual [118] | Reports missing or unbalanced delimiters and .EQ/.EN pairs | 1 |
| Col | 274 | Unix manual [118] | Filters reverse paper motions for nroff output for display on a terminal | 1 |
| Crypt | 121 | Unix manual [118] | Encrypts and decrypts a file using a user-supplied password | 1 |
| Tr | 127 | Unix manual [118] | Translates characters | 1 |
| TextureAtlas | 745 | GitHub [124] | Stores and manipulates multiple textures efficiently for graphics libraries | 1 |
| Chunkybar | *Unreported* | GitHub [124] | Implements multi-piece progress bars used in bittorrent clients | 1 |
| PseudoLRU | 384 | GitHub [124] | Implements the Least Recently Used (LRU) caching algorithm | 1 |
| QPHashMap | 1097 | GitHub [124] | Hashmap data structure that uses quadratic probing to manage collisions | 1 |
| ISSTA Containers | 2000 | ISSTA Artifacts [125] | Container classes | 1 |
| Java Collections 1.6 | 22000 | Oracle [126] | Java's collection library | 1 |

*Table II (Continued)*

| | | | | |
|---|---|---|---|---|
| ASM 3.1 | 40000 | Mavem repository [127] | A Java bytecode manipulation and analysis framework | 1 |
| Apache Ant 1.7.1 | 209000 | Apache [128] | A Java-based build tool | 1 |
| CCoinBox | 120 | Codeforge [129] | Library that simulates a vending machine | 1 |
| Calendar | 287 | Codeforge [129] | Library for calendar operation | 1 |
| Stack | 420 | Sourceforge [130] | Microsoft C# library for stack operation | 1 |
| Queue | 201 | Codeforge [129] | Microsoft C# library for queue operation | 1 |
| WindShieldWiper | 233 | Codeforge [129] | Library that simulates a windshield wiper | 1 |
| SATM | 197 | Codeforge [129] | Library that simulates an ATM | 1 |
| BinarySearchTree | 588 | Sourceforge [130] | Library for binary search tree algorithms | 1 |
| RabbitAndFox | 770 | Sourceforge [130] | Program that simulates a predator-prey model | 1 |
| WaveletLibrary | 2406 | Codeplex [131] | Library for wavel*et al*gorithms | 1 |
| BackTrack | 1051 | Sourceforge [130] | Library for backtracking algorithms | 1 |
| NSort | 1118 | Codeproject [132] | Library for sorting algorithms | 1 |
| SchoolManagement | 1726 | Sourceforge [130] | Program for managing school activities | 1 |
| EnterpriseManagement | 1357 | Sourceforge [130] | Program for managing enterprise business | 1 |
| ID3Manage | 4538 | Sourceforge [130] | Library for reading and writing of ID3 tags in MP3 files | 1 |
| IceChat | 71000 | Codeplex [131] | Program that implements an IRC (Internet Relay Chat) Client | 1 |
| CSPspEmu | 406808 | Github [124] | Program for a PSP (PlayStation Portable) emulator | 1 |
| Poco-1.4.4: Foundation | 149547 | Github [124] | Library that contains a platform abstraction layer and a large number of useful utility classes | 1 |
| Linear.ConjugateGradient | 107 | Commons [133] | An implementation of the conjugate gradient method | 1 |
| Linear.DefaultFieldMatrixChangingVisitor | 18 | Commons [133] | Default creation of custom visitors for matrix | 1 |
| Linear.EigenDecomposition | 344 | Commons [133] | Calculates the eigen decomposition of a real matrix | 1 |
| Analysis.function.Abs | 8 | Commons [133] | Absolute value function | 1 |
| Analysis.function.Gaussian | 81 | Commons [133] | Gaussian function | 1 |
| Analysis.function.HarmonicOscillator | 58 | Commons [133] | Harmonic oscillator function | 1 |
| Analysis.function.Sigmoid | 58 | Commons [133] | Sigmoid function | 1 |
| Analysis.function.Minus | 10 | Commons [133] | Minus function | 1 |
| Math.geometry | 340 | Commons [133] | 3D calculation | 1 |
| Math.util | 1161 | Commons [133] | Mathematic functions | 1 |
| lang | 4276 | Commons [133] | Basic utility | 1 |
| lang.text | 1475 | Commons [133] | Text processing | 1 |
| Collections.list | 823 | Commons [133] | A container structure | 1 |
| Action_Sequence | 2477 | EiffelBase [134] | Creates a sequence of actions to be performed on a call | 1 |
| Array | 1208 | EiffelBase [134] | Creates sequences of values of the same type which are accessible through integer indices | 1 |
| Arrayed_List | 2164 | EiffelBase [134] | Implements Lists using resizable arrays | 1 |
| Bounded_Stack | 779 | EiffelBase [134] | Implements Stacks with a bounded physical size, using arrays | 1 |
| Fixed_Tree | 1623 | EiffelBase [134] | Creates trees where each node has a fixed number of children | 1 |
| Hash_Table | 1791 | EiffelBase [134] | Stores and retrieves items identified by unique keys | 1 |
| Linked_List | 1893 | EiffelBase [134] | Creates a sequential, one-way linked lists | 1 |

*Table II (Continued)*

| String | 2980 | EiffelBase [134] | Sequences of 8-bit characters, accessible through integer indices | 1 |
|---|---|---|---|---|
| LockfreeList | 330 | [135] | typical lock-free concurrent list class | 1 |
| OptimisticList | 220 | [136] | A lock-based concurrent list class | 1 |
| LazyList | 200 | [137] | A lock-base concurrent list class | 1 |
| SimpleQueue | 93 | [138] | A simple locked-based concurrent queue class | 1 |
| RingQueue | 108 | GitHub [139] | A concurrent ring queue class | 1 |
| LockfreeQueue | 173 | [138] | A lock-free concurrent queue class | 1 |
| MSQueue | 160 | [138] | A lock-free concurrent queue class | 1 |
| BackoffStack | 205 | [140] | A concurrent stack class | 1 |
| LockBasedHashTable | 534 | GitHub [141] | A concurrent hash table class with striped locks | 1 |
| Bluetooth | Unreported | [31] | A Bluetooth application for file sharing | 1 |
| Contact | Unreported | [31] | Management of user contact | 1 |
| SMS | *Unreported* | [31] | SMS client for sending and receiving SMS | 1 |
| Bluetalk | *Unreported* | [31] | VOIP Bluetooth application | 1 |
| Dialer | *Unreported* | [31] | Make and answering calls | 1 |
| Browser | *Unreported* | [31] | Mobile Web Browser for surfing the Internet | 1 |

Therefore, the programs that are listed at the top of the table represent the most used among subject programs in the ART literature.

Out of the total number of ART studies considered for this review, only 38.5% (42 out of 109) have utilized real programs in their empirical evaluation. The results in Table II include subject programs from the ACM's collected algorithms (ACM CALGO) [114] and the Numerical Recipes book (Num. Recipes) [113], Software-artifact Infrastructure Repository (SIR) [115, 142], Siemens suite of programs (Siemens) [143] [115], Unix manual [118], GNU Scientific Library [116, 144], Apache Commons library (Commons) [133], EiffelBase library (EiffelBase) [134], ACM International Collegiate Programming Contest (in short as ACM-ICPC) programs [145], programs from Christopher Lott's website [119].

In total, we identified 123 different subject programs. A large number of them are written in C/C++ (53.6%) and Java (28.5%), with reported sizes that range from 8 to 406,808 lines of code. From the literature we obtained, the extremely large sized subject programs shown in the table were used by Chen *et al*. [72], where they tested various units (classes) of the programs with ART instead of testing the programs in entirety. Again, we found that the first 12 programs as shown in Table II, have been extensively applied by most ART studies that utilize real programs; all of which are programs from the Numerical Recipes book [113]. The 12 programs are simple real programs initially used by Chen *et al.* [27]. These programs involve numerical computations which were originally written in C, consist of between 30 to 200 lines of codes. From the table, the maximum number of times that a subject program has been used in empirical studies is 15, and there are a few programs with a high number of uses; whiles most of the programs have been used once.

### B. RQ2: VARIATIONS OF ART METHODS AND THEIR CHARACTERISTICS FOR NUMERIC PROGRAMS

ART is an enhanced form of RT, which improves the fault detection effectiveness of classical RT by imposing some additional criteria on the test inputs selection process. ART maximizes diversity by distributing the test cases evenly over the input domain [27, 76]. The evenly spread of random test cases over the whole input domain allows finding faults through fewer test cases than with purely random testing. As there are many possible ways to implement the concept of "even spread", a great number of ART methods have been proposed which provide diverse algorithms to address the "even spread" intuition. The various ART methods have different levels of performance depending on the cost of randomly generating an input, the cost of generating or selecting an input as a test case for that particular ART algorithm, the program execution time, the failure rate, or the failure patterns of the program under test. These methods attempt to maintain the benefits of random testing while increasing its effectiveness. These proposed ART methods achieve even spread either through a distance-based selection of candidates [27, 41], restriction of certain regions [34], partitioning of input domain [146], or other approaches that achieve diversity in input selection [40]. Empirically, studies [76], [31] show that the concept of even spread of the ART algorithms contributes to an improvement of 40% to 50% in failure-detection effectiveness over RT, which is close to the theoretical limit. Although ART has shown to be able to improve the fault detection effectiveness of RT it requires additional computational overhead in generating test cases, as larger previously executed test cases consequently reduce efficiency [60]. In this regard, researches into ART have yielded three overhead reduction strategies: *filtering* [147], *mirroring* [148] and *forgetting* [149], all of which reduce the computational costs associated with the ART algorithm. However, further researches to increase the efficiency ART have revealed other overhead reduction strategies.

Efforts to improve on the failure-detection effectiveness of ART and to reduce its overheads have resulted in the proposition of several ART methods. To answer the research question for this subsection (**RQ2**), we have classified the various ART methods that typically have notions to achieve an even spread of test cases, into distance/selection-based, exclusion-based, partition-based, and alternative-based methods, as illustrated in Figure 8. This classification is similar to the way we categorized the selected papers that propose ART methods for this review in Section III-E above. All ART methods have the characteristics of randomness in test cases selection and ensuring even spread of test cases within the input domain. The distance-based methods compute the maximum of all minimal distances between test case candidates and all previous test cases. The exclusion-based methods restrict regions in selecting test candidates as the next test cases. Although the exclusion-based methods utilize distance strategy, we classify them based on their definition of restricted regions within the input domain. The partition-based methods partition the area of the input domain and sample the next test cases from the partitions. While, the alternative-based methods employ strategies like [150, 151], that are based on different concepts from distance, exclusion, and partitioning in generating test cases.

As part of the requirements of the research question (**RQ2**), this section reviews ART methods under each category that have been proposed with a focus on their motivation and description.

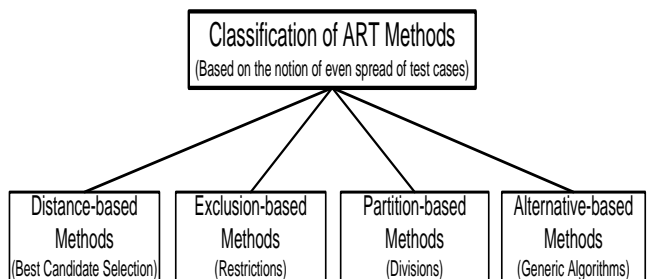

**FIGURE 8.** Classification of ART Methods.

### 1) DISTANCE-BASED ART METHODS
ART methods that fall under this category (also referred to as *Selection-based ART* methods), generally select candidate test cases and compute distances to or from them. Distance-based ART methods select the next test case from a set of candidates (candidate set generated randomly each time), by employing a distance criterion between the candidate cases and all the already executed test cases (i.e., test cases that have not revealed a failure).

The first and most extensively studied distance-based ART method, the *Fixed Size Candidate Set ART* (FSCS-ART), was proposed by Chen *et al.* [27, 41]. FSCS-ART was proposed based on Mak's original investigation into the strategy of maximizing minimum distances (*max-min* distance) among test cases [47]. Hence, FSCS-ART method is sometimes referred to as *Distance-based ART* (D-ART), as we have adopted for this category of ART methods. The method applies the basic ART algorithm as provided in Algorithm 1, and was motivated by the need to improve on the failure detection capability of RT. The algorithm of FSCS-ART selects test cases based on a *max-min* distance criterion or maximizing sum of distances (*max-sum* distance criterion) among test cases. The FSCS-ART process is illustrated in Figure 2 above

FSCS-ART algorithm has been commonly applied to numeric programs using the Euclidean distance metric, since finding the difference between similar primitive values is simple. To enhance the application of ART to other programs inputs—non-numerical inputs e.g., Boolean, strings, objects, arrays, structs, events, and other kinds of container data structures, etc—several researchers have applied the FSCS-ART algorithm with varying distance metrics. For instance, some works used the *category-choice metric* [42],[79],[66] to measure the distance between two inputs using their functionality triggers; *uncovered t-wise combinations distance metric* [105], [106] for determining dissimilarity for test inputs of the combinatorial test spaces (i.e., for inputs consisting of parameters, respective values, and constraints on value combinations); *static* and *dynamic metrics* [111] for concurrent data structures, and *coverage metric* [75], [78], [71] which utilize coverage information. On the other hand, Ciupa *et al.* [104], [28] introduced the *ARTOO metric* to compute distances between arbitrary objects in the algorithm of the ART. Despite its ability to support the full automation of applications in ART (e.g. AutoTest), the ARTOO metric has exponential calculation time (i.e., time complexity); hence, *Simplified Object Distance Metric*[108] and *Centroid-based metric* [73] were proposed to address the limitation. A generic distance metric that has also been applied to the original ART is the *Object and Method Invocation Sequence Similarity* (OMISS) metric, which uses semantic information (class information and input structure) to construct and calculate distances.

Although the original FSCS-ART is effective and has several variations, it has some limitations. For example, the computation in applying the distance criterion may be quite expensive, especially when the number of executed test cases increases [67]. Attempts to alleviate some pitfalls of the original FSCS-ART algorithm by employing various distance-based intuitions with varying motivations has contributed to the proposal of a number of ART methods in this category. The following are other Distance-based ART methods which are categorized based on similar motivations for their proposal.

*Distance-based ART Methods to Investigate and Improve on Effectiveness:*

Chen *et al*. introduced two additional versions distance-based ART [43]: Universal Candidate Set (UCS ART) and Growing Candidate Set (GCS-ART). The UCS ART method was focused on determining whether the difference in the F-measures between FSCS and RT was actually due to some discrepancies. The method discards a pool of randomly selected test candidates if all test cases in the pool are not failure-causing and selects a new test pool. Despite the fact that the exact same random test cases were used, UCS-ART had F-measures lower than those of RT. Improvements were generally smaller than with FSCS-ART. The GCS ART was introduced to make better use of the generation effort of FSCS-ART by retaining unselected candidates for later selection. The approach uses a candidate set that grows linearly in size as unselected candidates are kept for later rounds of selection, instead of discarding them. He found the effectiveness of this method to be a little inferior to FSCS-ART.

To enhance the spread of test cases more evenly than D-ART [47] in order to improve effectiveness, Mayer and Schneckenburger [152] introduced the *Continuous Distance-based ART* (C-D-ART). The test case generation algorithm of their approach was similar to that of D-ART apart from the way they computed distances. They modified the distance computation in D-ART such that all points within the input domain have virtually the same number of neighbors to achieve a better even spread of the test cases. Instead of computing distances just within the input domain, they regarded the input domain as virtually continuous, such that the opposite borders are directly adjacent to each other. Their approach provided better effectiveness than RT; however, its effectiveness depended on the dimensionality of the input domain.

An investigation to determine how the distribution of test cases correlates with the failure-detection effectiveness of an ART algorithm was carried out by Chen *et al*. [153]. In their work, they proposed an ART method based on *Distribution Metrics* (DM-ART), which primarily applies *discrepancy* and *dispersion* as selection criteria for test cases in ART. Discrepancy and dispersion are common metrics used for measuring the even distribution of sample points. A low discrepancy and low dispersion, not in isolation, provides an indication that sample points are to some extent evenly distributed, which implies an even spread of test cases [153]. These distribution metrics have not only been employed to measure and compare the even distribution of various ART algorithms but also they have recently been adopted as criteria for the generation of test cases with the aim of improving the even distribution of test cases and the fault detection capability of ART [37, 154]. Simulations and empirical results showed that the DM-ART not only improves the evenness of test case distribution but also enhances the failure detection capability of ART.

In a bid to investigate whether the application of Path Coverage information into ART can improve the effectiveness of test case selection process, Sinaga *et al*. [155] introduced the *Path Coverage for ART* (PC-ART) method. PC-ART computes the distance between test cases by using Path Coverage information (Branch coverage). From their experimental results, they concluded that path coverage information is good for defining distances between test cases in ART methods. In addition, PC-ART was found to be stable and significantly outperformed RT in terms of effectiveness.

*Distance-based ART Methods for Edge Preference Reduction:*

FSCS-ART has been found to have a bias of selecting test cases which are close to the edge of the input domain (referred to as *edge preference*), which is also referred to as the "*boundary effect*" [4], [42, 88, 156]. Such a bias may result in an uneven distribution of test cases, which affects the failure detection effective of the method. *Enhanced DART* [157] is a concept that reduces the boundary effect of distance-based ART methods by introducing virtual images of successful test cases around the input domain. The method uses the closest images (*effective images*) to the candidate test cases in distance computations. The method ensures that the preference of selecting test cases close to the boundary no longer exists in distance-based ART methods. Chen *et al*. [158] proposed a new approach, namely *ART with Dynamic Non-uniform Candidate distribution* (ART-DNC) to reduce the boundary effect of FSCS-ART. In their method namely FSCS-ART-DNC, the candidate case generation process was no longer conducted using a uniform distribution but was based on a dynamic non-uniform test profile. They selected one particular test profile and integrated such a profile with the test case identification process of the FSCS-ART algorithm. Their simulation studies showed that using the reduction of the preference for selecting test cases which are close to the edge of the input domain can significantly improve the effectiveness of the original ART algorithms.

Geng and Zhang [156] also introduced the algorithm of *Circling FSCS-ART* to enhance on the performance of FSCS-ART, as the algorithm of FSCS-ART was found to be more inclined to generate test cases on the boundary of the input domain. The algorithm of Circling-FSCS-ART mainly changes the bounded input domain into an unbounded one. Simulation results showed that the algorithm significantly reduces the boundary effect; thus improving failure detection effectiveness.

*Distance-based ART Methods to Improve Effectiveness in High Dimensional Input Domains:*

The work of Chen *et al*. [77] proves the probability of ART methods deteriorating in effectiveness for higher dimensions of the input domains. Some Distance-based ART methods have been proposed to mainly improve the fault-detection effectiveness of FSCS-ART for high dimensional input domains. Kuo *et al*. [58, 159] proposed an enhancement to FSCS-ART for high dimensional input domains. Their method applied FSCS-ART with an *eligibility filtering* process to ensure diversity of test cases in terms of parameters of the inputs. To use the method, the tester needs to set two parameters to control the eligibility criterion during the test process. The slight bottle-neck is the determination of appropriate parameters. The results of the simulation they performed showed that the method improves FSCS-ART in high dimensional cases, especially in situations that program failures depend on part of the input parameters. Kuo et al.[87] also proposed a variant of FSCS-ART, named *Inverted FSCS-ART*, which improves the fault detection effectiveness of FSCS-ART by inverting the edge/center distribution of FSCS-ART test cases. Their algorithm does not alter the test case selection process of FSCS-ART, but uses a linear function to map the selected test cases from the edge to the center region and vice versa before executing them. Their simulation results showed that inverting the test case distribution of FSCS-ART (which are normally edge-biased [77]) increases the chances of detecting the failure region.

*Distance-based ART Methods for Computational Cost Reduction:*

In an attempt to reduce the high distance computation requirement of FSCS-ART whiles maintaining wide spreading of test cases, Chan et al. [67] later introduced the *Centre-of-Gravity* (C.G.) constraint approach. Their method reduces the number of distance measures by introducing test candidates that are randomly generated from ART together with candidates from the C.G. constraint by a probability distribution. They realized that straightforwardly applying the C.G. constraint on the test candidates will result in black-hole effect [67], where future candidates that are chosen will be those close to the center-of-gravity of the input domain. Even though their approach aimed at improving computation cost, their empirical results also showed a slight improvement of 11% in failure finding effectiveness in addition to a significant improvement of 26% in computational cost than FSCS-ART.

Geng and Zhang [156] introduced a variant of the FSCS-ART algorithm named *Descending distance FSCS-ART* (D-FSCS-ART), which provided a reduced distance computation as compared to FSCS-ART. Their algorithm selects the next candidate test case by calculating the distance only on X-axis between the candidate test case and the elements at the index. Their simulation results showed that, without affecting the effectiveness, their approach reduces the number of distance computations in selecting the next test case.

2) EXCLUSION-BASED ART METHODS

This category of ART methods attempts to capitalize on the failure region contiguity, by using various intuitions that restrict certain regions of the input domain to achieve the ''even spreading'' of test cases. *Exclusion-based ART* methods achieve an even distribution of test cases by excluding regions close to previously executed test cases and randomly selecting new test cases away from these previously executed test cases. Although ART methods that fall within this classification also compute distances in their strategies, their classification as Exclusion-based is based on their basic use or definition of exclusion regions in their algorithms to attain diversity of test cases.

*Restricted Random Testing* (RRT) which was proposed by Chan et al. [33, 34] is the first exclusion-based ART method. The main motivation to its proposition, like any other ART method, was the intuition that, by incorporating additional information into the test case selection or generation process, it should be possible to improve the testing results of RT [160]. The method defines exclusion regions around all previous test inputs in the input domain and randomly generates test candidates one by one from the domain until a candidate falls outside of all excluded regions. This candidate is then selected as the next test case. All exclusion regions have an equal size, and the size decreases with successive test case executions. The exclusion regions normally have the shape of either a circular or square for a 2-dimensional input domain [34]. They confirmed the correctness of the intuition with experimental results – in the experiments, the RRT method significantly outperformed RT (on some occasions, by up to 80%) [34]. Although the use of simpler exclusion regions (e.g. squares) would make the calculation of exclusion relatively trivial, it was found that the best failure-detection performance was obtained when circular regions were used [34]. Therefore, exclusion algorithms normally calculate using circular exclusion regions.

Various researchers have attempted to employ several intuitions of exclusion with varying motivations, which has resulted in the proposition of some variations of Exclusion-based ART methods. Below are descriptions of other Exclusion-based ART methods, which are categorized by the similarity of motivations for their proposal.

*Exclusion-based ART Methods to Investigate and Improve on Effectiveness:*

To enhance the spread of test cases more evenly than RRT in order to improve effectiveness, Mayer, and Schneckenburger [152] also introduced the *Continuous RRT* (C-RRT), apart from their C-D-ART discussed above. Their approach was similar to that of RRT apart from the way they computed distances. They modified the distance computation in RRT such that all points within the input domain have virtually the same number of neighbors to achieve a better even spread of the test cases. In computing distances within the input domain, they regard the input domain as virtually continuous such that opposite borders are directly adjacent to each other. Their approach provides better effectiveness than RT; but, its effectiveness depends on the dimensionality of the input domain (i.e. its effectiveness may be affected by certain irregular dimensions of input domains).

*Exclusion-based ART Methods for Edge Preference Reduction:*

For RRT, candidates close to the boundary of the input domain have a higher chance to be outside all exclusion regions than those close to the center. *Enhanced RRT* [157] is an innovative concept that reduces the boundary selection candidate tests of RRT by introducing virtual images of successful test cases. Its algorithm only checks whether the candidate is outside the exclusion region of the closest image. The method reduces the boundary selection of test cases. Similar to ART-DNC discussed above, Chen *et al.* [158] proposed a variant of the RRT algorithm and used it to reduce the edge preference of test case selections. Their investigation of the frequency distribution of tests generated by RRT, showed that points located at the boundaries of the input domain had higher probabilities of being selected as test cases than those around or close to the center of the input domain. Their proposed method RRT-DNC also generates test candidates using a dynamic non-uniform test profile. They integrated their test profile with the test case identification process of the RRT algorithm. Their algorithm reduced the edge preference in test case generation and showed better fault detection capabilities in contrast with the original RRT. The

*Exclusion-based ART Methods for Non-homogeneous Inputs:*

As RRT imposes a uniform exclusion zone that is centered on each executed test case (eg., circular in 2D), its application to programs whose input domains are less homogeneous (eg., not square in 2D), may result in an unanticipated bias of the exclusion region. In addition, the value of Max R (the optimal value for the Target Exclusion Ratio) is not easily estimated when the input domain is considerably disproportionate. Chan *et al.* [161] incorporated a normalizing feature into the RRT to produce the *Normalized RRT* (NRRT) Method. The method defines an exclusion zone that is scaled to the shape of the non-homogeneous input domain around each non-failure-causing test case. NRRT alleviated the potential difficulty in defining exclusion regions within non-homogeneous input domains and facilitated the prediction of Max R.

*Exclusion-based ART Methods for Rigidity Reduction to RRT:*

Some researches [162, 163] argued on the rigidity of RRT in discarding all test cases within the exclusion regions. They argued that some of the discarded test cases may detect software failures. *ART by Exclusion through Test Profiles* (ART-E$_{TP}$) was proposed by Liu *et al.* [162, 164] to minimize the rigidity of RRT in discarding all test cases within exclusion regions. ART-E$_{TP}$ simply selects test cases using a well-designed test profile. The method assigns a zero probability value to all executed test cases; but assigns certain degrees of probability values to all other points that are potential candidate test cases, according to their proximity to the executed test cases. From the experiments they performed [162], it was evident that ART-E$_{TP}$ provided reduced computation overhead as compared to RRT, although the method did not omit any program input that may be failure-causing. Their results also showed higher performance in failure-detection effectiveness, as the method spread test cases more evenly. A major disadvantage of this method is that it exhibits an uneven distribution of test case for high dimensions of input domain. In addition, it exhibits better performance only with a single test profile. Chan *et al.* [163] also introduced the *Probabilistic ART* (PART) method aimed at minimizing the rigid or strict exclusion of RRT. Their approach included all regions of the input domain in the generation of test cases, but with a bias. They generated test cases even within restricted regions but with proportionately less probability of being selected than test cases further away. Though their method achieved the generation of test cases from all regions of the input domain, its performance was not very impressive as compared to RRT. The method requires further investigations, especially pertaining to the choice of values for its control parameters [163].

*Exclusion-based ART Methods for Computational Cost Reduction:*

Though RRT provides very high failure-detection effectiveness in terms of F-measure, it incurs some computational costs in its restriction algorithm [85], [149]. For each acceptable candidate test case, it is possible that several attempts at generating test case outside an exclusion region will have failed. Chan *et al.* [147] introduced an ART method that employs the *filtering* strategy namely *ART with filtering*, to reduce the overheads associated with the generation of an acceptable test case while maintaining the failure-detection effectiveness of the basic ART methods. They defined a bounding region around a candidate test case to filter the previously executed test cases and calculated the distances only from the candidate to those executed test cases inside the bounding region. Their method attained significantly reduced computational overheads as compared to most basic ART methods. Although this filtering approach employs distance computation, the definition of a bounding region to restrict some test cases formed the basis for its classification as Exclusion-based ART method. Chan *et al.* [149] also proposed the *RRT with forgetting* strategy to retain much or all of the failure-revealing effectiveness in addition to the reduction of overhead costs of RRT. Instead of utilizing all previously executed test cases in the test case generation algorithm, forgetting evaluates only a subset of the executed set when selecting the best candidate. They proposed three forgetting schemes: *Random forgetting*; *Consecutive retention*; and *Restarting*. All three schemes of the forgetting strategy reduce computational overheads; however, they do not consider the locations of candidates and test cases. Therefore the wrong choice of test cases for the limited number of executed test set in the restriction algorithm may adversely affect its performance on some subject programs. Moreover, forgetting information about some previously executed test cases in test cases selection slightly deviates from the main idea of the ART technique [27, 76].

A very recent work by Ackah-Arthur *et al.* [165] that adopts the notion of exclusion, was aimed at minimizing the computation overhead of ART methods. They constructed a new version of ART that considers the distribution of test candidates spatially. Their method referred to as the *Candidate Exclusion ART (CE-ART)*, provides flexible exclusion zones around candidate test cases. Their method restricts distance computations from a candidate to only previously executed test cases that reside within its exclusion zone in order to select the farthest among them. The method also selects candidates with empty exclusion zones directly as inputs. Their experimental results indicated that CE-ART provides significantly low computation overheads whiles maintaining comparable failure detection effectiveness.

### 3) PARTITION-BASED ART METHODS

ART by partitioning is another typical notion to evenly spread test cases and is inspired by *partition testing* [54]. Methods in this category referred to as *Partition-based ART* methods use a rather different intuition—in the sense that, they partition the input domain and sample from partitions obtained to achieve even spread of test cases; therefore they employ the strategy of Proportional Sampling [56]. In generating test cases from respective partitioned regions, Partition-based ART methods can either randomly select [146], or use a distance-based ART algorithm (eg. FSCS) [39, 154, 166], or use an exclusion-based ART algorithm (eg. RRT) [35, 37, 167]. Therefore this category of ART methods can also be referred to as the "hybrid" category, as they mostly involve methods that combine other notions in order to increase performance. Even though ART methods that fall in this classification can combine other strategies like the distance, and exclusion strategies in their algorithms, their classification as Partition-based methods is however based on the use of partitioning in their algorithms to attain diversity of test cases.

In 2004, Chen *et al.* [146] proposed the first Partition-based ART, namely *ART through Dynamic Partitioning (ART-DP)*. Since then there has been an explosion of work, applying partitioning strategies in a wide range of ART approaches. The strategy of Chen *et al.* [146] was part of efforts to reduce the computational time of ordinary ART in order not to outweigh the advantage of performing fewer tests; especially in situations where the cost of test selection is less. They defined two main schemes of methods: *ART by Bisection* (ART-B) and *ART by Random Partitioning* (ART-RP). ART-B bisects the longest coordinate of the input domain to obtain equally-sized partitions. The method randomly generates a test case from each partition and iteratively subdivides all partitions into halves once all partitions contain test cases. On the other hand, ART-RP samples a test case based on the proportional size of region area to the whole input space. The method iteratively partitions the input domain using the previously executed test cases themselves (that is, dividing a region by drawing perpendicular lines on the location of the most recently executed test case within the input space) and then chooses the maximum-sized region to randomly generate the next test case. Although it avoids distance computations, ART-RP has to search through the whole input domain for the largest sub-domain to generate the next test case, and this presents a slight overhead especially when the number of sub-domains increases. Generally, as ART-DP algorithms do not involve distance computations and many comparisons, their time costs are low compared to those of other ART approaches; but at the cost of lower failure detection capabilities, especially for strip and point failure patterns. Attempts to utilize the continuity of failure regions, by employing several intuitions of partitioning with different motivations, has led to the proposition of other Partition-based ART methods by various researchers. These proposed methods normally provide improvements to the schemes introduced by ART-DP. The following are descriptions of the proposed Partition-based ART methods which are categorized based on similar motivations for their proposal.

*Partition-based ART Methods for Enhancing the Effectiveness of ART-DP:*

Some Partition-based ART methods were mainly proposed to improve upon the first Partition-based ART algorithms, in terms of failure-detection effectiveness. Such methods include *ART by Bisection and Localization*, and *ART by Bisection with Restriction*, both of which were introduced by Mayer [168]. His motivation was to provide a modification of the ART-B that can improve its failure-detection effectiveness without affecting its computational efficiency. The algorithm of *ART by Bisection and Localization* combined ART by Bisection with the principle of localization using either D-ART or RRT. The algorithm selects the next test case from "empty" regions of the partition using D-ART or RRT, where it only performs the distance computations with "neighboring" previously executed test cases. When RRT is used, the algorithm rejects a randomly chosen test candidate if the distance to one of its neighboring regions is not greater than the radius of exclusion. When D-ART is used, the algorithm selects fixed-sized test candidates from an "empty" region of the partition and chooses the candidate with the maximum of all minimal distances to neighboring previously executed test cases as the next test case. Although his algorithm required only a linear number of distance computations in order to detect a fault, the method significantly improved the failure detection effectiveness of ART-DP. On the other hand, the *ART by Bisection with Restriction* algorithm combined ART by Bisection [146] with the notion of homogeneous restriction [33, 34]. Using the idea of restriction, the algorithm begins as ART by Bisection, except that it also selects test cases from homogeneous restricted sub-domains. The algorithm is simple and it does not require any distance computation at all in its test case selection process. Results of simulations showed that the algorithm provides improved failure-detection effectiveness as compared to ART-B. Also, its effectiveness can be compared to D-ART and RRT. Another partition-based approach to improving the failure detection capability of ART-B is the *ART-B by Flexible Partitioning strategy and Candidate Strategy* (B-ART-FPCS), which was introduced by Mao and Zhan [169]. In the B-ART-FPCS algorithm, two partitioning strategies were implemented: (1) *Flexible partitioning strategy*, designed to extend the splitting line (or plane) farther away from the test case within the region being bisected. (2) *Candidate strategy*, designed to select an appropriate candidate that has a greater boundary distance as the next test case. Although their method presented high test case generation time due to the additional computation about splitting coefficients of dimensions and boundary distances of candidates, their simulation analysis confirmed a linear-order time complexity of BART-FPCS. From the results of their simulation analysis, the B-ART-FPCS algorithm exhibited stronger failure detection effectiveness than the original ART-B algorithm in most cases.

Mayer [35] later proposed another enhancement method particularly to improve the failure-detection effectiveness of ART-RP (a method scheme of ART-DP), using the strategy of restriction. The method, referred to as *Restricted ART by Random Partitioning*, iteratively subdivides the largest region of the input domain using the location of a newly generated test case. Instead of selecting test cases from sub-regions of the input domain just like in ART-RP, the method rather chooses from restricted versions of these regions. His simulation study showed that the method can better detect failure and presents similar runtime as compared to ART-RP. Another effective improvement to ART-DP is *ART by Two-Point Partitioning* (ART-TPP) proposed by Mao [36]. His method partitions the current maximum-sized region using the midpoint of two previous test case locations (points) in the region instead of on a single point (as employed by ART-RP). The method first generates a point randomly, and then picks out the second point from a candidate set using the farthest distance criterion. Simulation analysis performed showed that ART-TPP was generally better in failure detection effectiveness than ART-RP but worse than ART-B, although its algorithm was more stable. From our evaluations of the study, further research must be performed to validate its findings as the work considered only two-dimensional input domain to perform its simulation analysis.

*Partition-based ART Methods to Eliminate Closeness of Test cases in ART-DP:*

Liu *et al.* [164] argued about the limitations of ART-DP algorithms, as they highlighted the possibilities of generating test cases that are close to each other. They, however, introduced two methods that improve on the limitations of the different ART-DP method schemes using a well-designed test profile. They proposed the *ART by Random Partitioning through Test Profiles* (ART-$RP_{TP}$) to improve the limitations in test case selections of ART-RP. In their algorithm, the farther away a point within the largest partition is from the already executed test cases, the more priority it has to be selected as a test case, and the points inside all other partitions should have no chance to be selected as test cases. They also proposed the *ART by Bisection through Test Profiles* (ART-$B_{TP}$) to similarly improve on the limitations of test case selections of ART- B. In their algorithm, the farther away a point inside an empty partition is from the boundaries, the more priority it has to be selected as a test case; and, all points inside non-empty partitions have no chance to be selected as test cases. The use of priorities ensured that generated test cases were not close to each other. Experimental results showed that the approaches of Liu *et al.* [164] provide higher failure-detection capability than both ART-RP and ART-B due to the elimination of their test case closeness limitations.

*Partition-based ART Methods for Computational Cost Reduction:*

Other partition-based approaches have also been introduced that seek to mainly improve the computational cost of ART while maintaining comparable failure-detection effectiveness to the classical ART algorithm. Chen *et al.* [167] introduced one of such approaches referred to as *ART by localization*. Their method localizes the test case generation region and previous test cases within the region and then generates the next test case from the restricted test case generation region using D-ART or RRT. However, their method had the limitation of identifying previous test cases located on the vertices of test case region as nearby executed test cases; some previous test cases may also be close to the edges of the test case generation region. Empirical results showed that in addition to attaining a comparable fault detection capability to those of other ART methods, their method showed a more reduced computational cost. *ART through Iterative Partitioning* (IP-ART) [49] is another method proposed to reduce computational costs. IP-ART uses partitioning to identify a region that is far apart from all previously executed test cases, to generate the next test case. IP-ART generates totally new and finer partitions of the entire input domain when no distant partitions are available to generate test cases from. In the algorithm of IP-ART, the overheads in terms of the cost of repartitioning the input domain and location of regions within the partitions are reduced, as it divides the whole input domain into equally sized grid cells and locates regions using the grid coordinates; thus significantly reducing its time complexity. Chow *et al.* [37] also proposed the *ART of Divide and Conquer*, which bisects the input domain into smaller equal sized sub-domains. Their algorithm then generates the same number of test cases inside each sub-domain; while using a *threshold* to limit the computational growth when a large number of previously executed test cases are involved. However, the limitation of their method was the difficulty in determining the threshold. Although the approach provided improved computational costs, its efficiency depended on the value of the threshold. On one hand, the algorithm does not significantly reduce the computational cost is not significantly reduced if the threshold is set to be too high. On the other hand, if the threshold value is too small, the overall process will result in the domination of the input domain by many sub-divisions. Sabor and Thiel [170] also introduced the *ART by Static Partitioning*, which partitions the input domain into equilateral cells and uses coloring techniques to color cells. With the use of the coloring technique, the algorithm is able to select test cases from cells which have the farthest distance to the cells which already contain previously executed test cases. Empirical results showed that in addition to providing similar effectiveness performance to existing ART methods, their method achieved far less computational overhead (linear time complexity).

Another partition-based ART method that was introduced mainly to reduce the computational overhead of the classical ART algorithm is *ART by Distance-aware Forgetting* (DF-FSCS), which was introduced by Mao et al. [39]. DF-FSCS minimized the limitation of the original forgetting strategy [149] by considering the geometric locations of forgotten test cases. In their algorithm, a given candidate ignores test cases out of its "sight" and computes distances only to neighboring test cases, using the spatial distribution of the test cases. The algorithm dynamically adjusts partitions and applies second-round *forgetting* [149] to ensure linear complexity. It significantly lowers the computational overhead of common ART algorithms such as FSCS and RRT while largely maintaining similar effectiveness. Though DF-FSCS has low computational cost, Rezaalipour *et al.* [171] have reported its limitations—generating test cases in highly populated areas and defining smaller regions at the lower or upper boundaries of input domains—which affects its even spread of test cases. They propose an improvement to DF-FSCS named *Arselda*, which generates candidate test cases from cells with the least number of test cases and enlarges the boundary regions to increase the quality of test cases generated. According to their experimental results, Arselda shows improved failure detection effectiveness for the block failure pattern and attains lower computational overhead than DF-FSCS.

Research by Chan *et al.* [148] to reduce the computational cost that is associated with the algorithm of RRT yielded the *mirroring* strategy. They applied the mirroring strategy to ART, by partitioning an input domain using a partitioning scheme (also known as *mirror partitioning*), applying the RRT method to only one of the sub-domains, and mapping the test cases (either by *reflection* or *translation*) to the other sub-domains. Since a method that employs the *mirroring* strategy partitions the input domain and selects test cases from individual sub-domains of the input domain to cause even spread, we classified such a method as part of Partition-based ART methods (see Figure 8 above). Their method proved to be a very attractive variation of RRT as it very much alleviated computational costs without affecting failure detection performance of the general algorithm of RRT in test case generation. A potential application of *Mirroring* is in those situations where the input domain is not regular, as applying RRT in less regularly shaped input domains is problematic [86]. With *Mirroring*, it should be possible to partition such input domains to create a relatively regularly-shaped source subdomain. Chen et al. [172] similarly applied the mirroring strategy to ART using the FSCS-ART algorithm to generate test cases in each partition. They introduced the *Mirror ART* (MART) to reduce the computational cost of the FSCS-ART method. The results of their simulation proved that MART does improve the cost-effectiveness of ART. To further minimize the order of computational overhead for MART, Huang *et al.* [173] varied the mirror function in their *Dynamic Mirror ART* (DMART) method. Their algorithm incrementally partitions the input domain and adopts new mirror functions that are dynamic (and not static) to provide higher efficiency. Experimental results confirmed its reduction in computational cost as compared to MART.

*Partition-based ART Methods for Edge Preference Reduction:*

As mentioned earlier in this study, studies conducted to compare ART methods [42, 79, 88] revealed that some of the methods have a preference of selecting test cases from close to the edges of the input domain (edge preference) over from the center; thus affecting the performance of these methods under some situations like increase in failure rate and dimension. To reduce the edge preference in selecting test cases from the input domain over the center, some Partition-based ART methods have been proposed. Two such methods were introduced by Chen *et al.* [154, 166] to reduce the edge preference problem of FSCS-ART and RRT respectively. They first proposed the *FSCS-ART with Partitioning by Edge and Center*, which partitions the input domain from the edge to the center. The algorithm generates a set of candidate test inputs from the input domain and if any of them is inside the same partition as a previous test case, it is randomly replaced. It then selects the farthest candidate test input from previous test cases as the next test case. Their second method, *RRT with Partitioning by Edge and Center*, also partitions the input domain from the edge to the center. It defines an exclusion zone around previous test cases and randomly generates inputs. A test input is selected as the next test case if it is outside both the exclusion zone and the partition of previous test cases. From a series of simulations they conducted, the two methods distributed test cases more evenly and provided improved failure-detection capabilities as compared to FSCS-ART and RRT, due to their reduced edge preference in the selection of test cases. A later work by Sabor and Mohsenzadeh [38] also resulted in an enlarged input domain approach that decreases the edge preference problem in most partition-based ART methods like *ART by localization* [167]. Their approach referred to as *ART through Dynamic Partitioning by Localization with Restriction and Enlarged Input Domain*, initially enlarges the input domain by a factor. It then selects test cases from the enlarged input domain and uses them in partitioning the original input domain; however, the selected test cases are executed only if they exist in the original input domain. Simulation results indicated a reduction in the preference of generating test cases from the edge of the input domain. The results also showed that the reduction of the edge preference can increase the performance of ART.

*Partition-based ART Methods to Improve Effectiveness in High Dimensional Input Domains:*

From the work of Chen *et al.* [77], there is the likelihood of some ART methods deteriorating in effectiveness as the number of dimensions of the input domain increases. Therefore, Chen *et al.* [174] introduced a Partition-based ART method, referred to as *ART by Balancing*, with the main aim of improving the fault-detection effectiveness of ART for high dimensional input domains. Their method ensured that the centroid of test cases in each partition of the input domain was close to the centroid of the corresponding partition, and applied a *Balancing* strategy to select test cases. Simulation results showed that the fault-detection capability of ART by balancing outperforms other ART methods greatly in high dimensional input domains.

4) ALTERNATIVE-BASED ART METHODS

ART methods that fall into this category employ varying diversity algorithms which are different from the notions of the distance-based, exclusion-based and partition-based methods but obey the principles of ART to achieve even spread in test case generation. We refer to the methods in this category as *Alternative-based ART* methods in the rest of this study. This subsection summarises some proposed Alternative-based ART methods obtained from the literature we gathered for this study. Below are descriptions of proposed Alternative-based ART methods which are categorized according to the commonality of motivations for their proposal.

*Alternative-based ART Methods for Improving Effectiveness:*

Some Alternative-based ART methods were introduced to minimize the number of test cases required to find a failure to improve the effectiveness of ART over RT. One of such methods is *Fuzzy ART* (FART), which was proposed by Chan *et al.* [151]. Their approach uses *fuzzy reasoning* [175] to guide the selection of test cases by evaluating potential test cases. FART was proven to significantly reduce the number of test cases required to detect failure. Moreover, Tappenden and Miller [150] also employed an evolutionary search algorithm to increase the effectiveness of ART by making efforts to maximize the test coverage of the input domain. In their method, named *Evolutionary ART* (eART), the evolutionary search algorithm was used to find an approximation for the test case that has the maximum distance from all previous test cases in its selection process. The results of an extensive simulation analysis found the evolutionary approach to be superior in effectiveness as compared to RT and other ART methods, especially amongst block failure patterns. In addition, they found that its application is feasible, and within the same order of time complexity as the other ART approaches. The main disadvantage in their work is the painstaking in the selection of an appropriate fitness function, and possibly functions that are tailored for specific failure patterns and input domains.

*Lattice-based ART* (L-ART) provides a further advancement to ART. L-ART is a distinctive ART method introduced by Mayer [176] to select high-density test cases from the input domain. In generating test cases, his algorithm systematically places test inputs to maximize the distances between them, and then randomly shifts lattice nodes in the input domain to further increase diversity. Apart from having a better fault-detection capability than RT, the algorithm has a very good performance than that of common ART algorithms. However, L-ART may have its generated test cases to be highly concentrated on some parts of the input domain and can cause a skewed distribution of test cases. This skewed distribution of test cases can result in a tight coupling between the fault detection capability and the location of the failure region in the input domain. This means when failure regions, unfortunately, reside away from the area where L-ART selects a high density of test cases, the algorithm may show a worse fault detection capability as compared to when failure regions are in the high-density area. To take care of the possible skewed distribution of test cases, the L-ART algorithm was further enhanced by Chen *et al.* [177] with the proposal of the *Enhanced Lattice ART*. The enhanced algorithm divides the input domain into equally sized grid cells and strictly restricts the locations of test cases whiles generating new test cases only from regions that have not been occupied by successful test cases. Investigations showed that this enhanced algorithm reduces the skewed test case distribution and provides a better and more consistent fault-detection capability (effectiveness) than the original L-ART and even other ART methods [35, 48, 49, 168, 174, 178].

Whiles other ART methods guided the distribution of test cases [28, 167] to improve on the effectiveness of RT, *Path-sensitive ART* (PathART) which was proposed by Hou *et al.* [145], rather utilized the program-path information to improve the effectiveness of RT. Their approach randomly generates test candidates and evaluates their deviation from some path constraints. And then selects as the next test case, the candidate that is farthest away from the others according to their path constraints. Experimental results showed that PathART improves the effectiveness of RT and aims to generate test cases evenly distributed on different execution paths of the program under test. However, their method is computationally expensive, as the algorithm has to go through the process of analyzing execution paths of the program, determining constraints for executing the paths, calculating the path distance between test cases according to their satisfaction for paths' constraints, and finally generating test cases far away from each other according to their path distance.

Hui and Huang [179] adopted a metamorphic distance metric [180] into ART to generate test oracle that can easily be validated for its correctness. They referred to their method as *Metamorphic Distance-based ART* (MD-ART). They introduced the *Metamorphic Distance* to compute distances between candidate test cases in order to make the test cases distribute in the input domain as widely as possible. The results of their primary experiment showed that MD-ART generated more effective test cases than RT. However, from the perspective of algorithm complexity, the complexity of MD-ART was higher than previous ART algorithms.

*Alternative-based ART Methods to Improve Effectiveness in High Dimensional Input Domains:*

Schneckenburger and Schweiggert developed the *Search-based ART* (SB-ART) [40] method to improve on the low effectiveness of ART approaches in higher-dimensional input domains since in a reality the testing area of the input domains is usually far from being one- or two-dimensional. Their method was based on the idea of the local search technique '*Hill Climbing*' [181]. In their approach, an initial test case set obtained from any ART approach (e.g. D-ART) is shaken as long as their fitness–measured as the shortest of all distances between any two test cases–increases articulately. The scaled fitness of the resulting test set slightly increases for a higher dimension. From experiments performed, SB-ART provided very little enhancements in effectiveness to D-ART. The SB-ART did not fully solve the dimensionality problem since its algorithm required prior knowledge of the failure pattern geometry, which is generally not available at the beginning of a search.

*Alternative-based ART Methods for Computational Cost Reduction:*

Although most Alternative-based ART methods discussed were proposed to improve effectiveness, *ART by Voronoi Diagram* was introduced by Chen et al. [178] mainly to improve the test case generation overheads of ART. The method uses a geometric data structure to reduce overhead and provide further optimization. They demonstrated that ART implementation using the Voronoi diagram data structure can be a more computationally efficient approach to generate test cases than other ART methods.

Another ART approach to reducing costs of generating test cases employs the Voronoi diagram data structure, namely *Random Border Centroidal Voronoi Tessellations* (RT-RBCVT). The RT-RBCVT algorithm proposed by Shahbazi et al. [61], is an innovative linear-order test generation method for numeric program inputs that use the centroids of Voronoi regions and a probabilistic method to produce an improved set of test cases. They then proposed an optimized RT-RBCVT calculation method (RT-RBCVT-Fast) that employed a search algorithm to generate test cases with a linear runtime. Their RT-RBCVT-Fast method is of the same order of computational complexity as RT. This significantly provided evidence that ART can indeed serve as a cost-effective alternative to RT. Though both RBCVT and RBCVT-Fast algorithms provide significantly reduced computational costs, they are limited by the need to determine a specific number of test cases for testing at the beginning. The problem is the difficulty in assuming the size of test cases (over-estimation or under-estimation), especially in the lack of information about failure rate, can affect the stability of its failure-detection capability.

### C. RQ3: TRENDS OF CONTRIBUTIONS AND DEVELOPMENT IN ART METHODS FOR NUMERIC PROGRAMS

Among the papers considered in this study, we identified 60 varying ART methods. The study has categorized the ART methods into four, as discussed in Section IV-B. We compared the total number of ART methods under each category, which represent the contributions to each category of methods. Figure 9 below shows the results of the comparison.

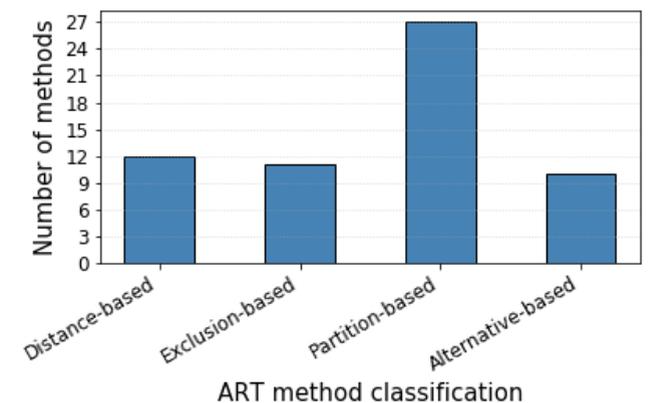

FIGURE 9. Contributions to each Category of ART Methods.

The results in Figure 9 indicate that Partition-based category has the most ART methods whiles the contributions to distance-based, Exclusion-based, and Alternative-based methods are similar though relatively high.

We then evaluated the yearly study contributions of the individual categories of ART methods, in order to ascertain their development trend over the years from 2001. Figure 10 below illustrates the individual trend of results obtained.

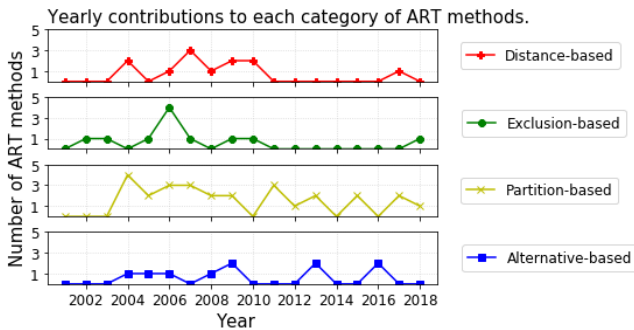

FIGURE 10. **Yearly Contributions of the Individual Categories of ART Methods.**

Comparably, it is obvious that there has been far less interest in developing ART methods that solely employ either the notion of Distance or Exclusion methods in recent years; whiles the Alternative-based ART methods have seen consistently low yearly contributions over the years. The contributions to Partition-based methods have been relatively high since the year 2004. In the year 2004, the contributions to the Partition-based ART methods reached an all-time high level as compared to the other categories of ART methods.

In order to further investigate the general trend in the number of ART method contributions, we compared the total number of ART methods proposed each year. The summary of the comparison from the year 2001 to 2018 is shown in Figure 11.

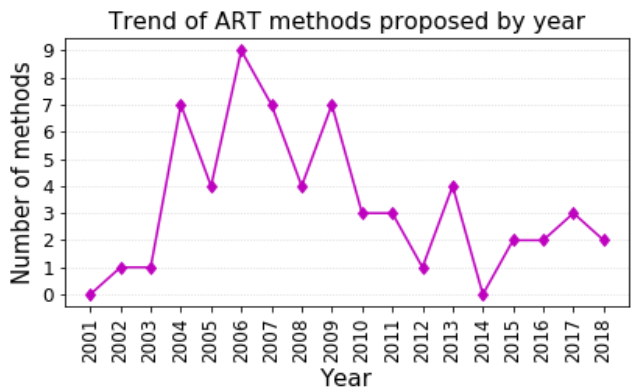

FIGURE 11. **Yearly Number of Contributions to ART Methods (2001-2018).**

The number of contributions to ART methods reached a record maximum in the year 2006. This performance has remained consistent to date with three ART methods proposed on the average for every year, except for the year 2014 which is unique which is unique as no method was proposed.

## V. DISCUSSION

In this section, we discuss some of the findings from the review results obtained in Section IV to provide answers to the research questions *RQ1* to *RQ3*. We then discuss some open issues related to ART and provide some future directions to the research field. Lastly, we present the limitations to the study. Naturally, this account is, to some extent, influenced by the authors' own position on ART. However, we have discussed the findings in an objective manner, based on the available literature and the trends and contributions among them.

### A. FINDINGS AND CURRENT RESEARCH TRENDS AND CONTRIBUTIONS

#### 1) RQ1–TRENDS AND CONTRIBUTIONS OF RESEARCHES IN ART

From the evaluation of the 109 ART papers selected from 2001 to early 2018, it is clear (see Figure 4) that significant research efforts have been made towards each category of ART studies. From Figure 4, it is evident that many (49.5%) studies on the theory of ART have been done as compared to studies that propose ART methods. Among the identified studies that propose ART methods, studies that employ the partition strategy in ART are more (23.9%) than those that employ the distance, exclusion, and other strategies in ART for test case generation.

The results in Figure 5 indicate that studies in ART have been quite unstable but with consistent contributions from 2001 to 2018 according to the number of yearly ART studies. The yearly studies can be classified into three stages. The first stage was from 2001 to 2003. In this early stage in which the idea of ART was first proposed, several simulations were performed and used to support the establishment of the fundamental theory of ART, such as the Failure Distribution [41, 77] and Validity [47, 83]; thus heightening researcher interests in the subsequent stage. The second stage, which was between 2004 and 2011, saw increases in the number of ART studies representing a high interest in the field. In 2006, for the first time, the total number of researches on ART peaked. From the evaluation of the studies presented at conferences and workshops (represent 63.3% of the total studies retrieved), it is possible that the reason for the spike in studies will have been influenced by the number of ART-related conferences and workshops held in those years. That is because there were as much as 19 software engineering related conferences and workshops held between the year 2004 and 2006 compared to 16 related conferences from the year 2007 to 2009 and 9 related conferences from the year 2010 to 2012. This heightened study contributions gradually declined till the year 2011. We think that the decline in the number of ART studies within the period may be due to a number of factors; such as possible limitations[104] in applying the technique to software that takes other inputs apart from numeric. Moreover, the yearly numbers of studies for the period were relatively significant. The third stage was from 2012 to 2018. It is obvious from Figure 5 that the number of research work on ART has been quite stable over the period, with consistent contributions of averagely five publications each year. Analysis of the works within this period provided compelling evidence that most studies of ART focus on developing strategies (1) that improve the efficiency of distance computations in ART [39, 59, 66, 70-73, 155, 165]; and (2) that are applicable in other testing scenarios [7, 97, 98, 102, 107, 165, 179]. Hence, the focus is on computational efficiency and application in other test settings. Figure 5 clearly shows an improvement in the number of ART studies in 2017 and the values of 2018 suggest a possible sign of further studies in the field.

As discussed earlier in the study, the *Analysis* category of studies consists of all studies that do not propose any ART methods but provide either meta-studies about ART, hypotheses supporting ART, evaluation of ART features, criticisms of ART, distance metric approaches, theses, reviews, and others. The *Analysis* category has gained the highest number of study contributions over the years. The higher number of yearly contributions to the *Analysis* category of studies, as shown in Figure 6, provides evidence that there have been relatively significant studies on the theory of ART.

Among all the studies identified that propose ART methods, the partition category of ART studies has experienced generally very high contributions over the years. However, fewer study contributions have been made towards both ART studies that use the distance notion, and those that use other notions different from distance, exclusion, or partition (as indicated in Figure 6 with *Alternative* category classification), to propose ART methods. The slight increase in contributions to the Alternative category over the years may be that researchers have found effective other strategies that can also ensure the random and even spread of test cases. Studies that propose ART methods using the exclusion notion have generally received the lowest contributions over the years although it seemed very promising in the year 2006. From our evaluation of all studies, the exclusion notion has mostly been integrated into other variations of ART method in other studies [168], [146], [35], [37], [167]. This indicates that improved ART performance is achieved when the exclusion notion is integrated into other notions, especially the Partition notion. Hence, the lower contributions to the exclusion category of ART studies over the years. Generally, the results in Figure 6 are an indication that averagely, most studies in ART have focused on theoretical issues about ART apart from the numerous studies that proposed methods for ART. This provides an obvious interpretation that ART has experienced much more in-depth studies. The results of studies in Figure 6 are similar to those obtained in Figure 4.

Researches in the area of ART appear to be very diverse as most of the authors who have contributed to the area have authored or co-authored in one paper, with few of them dominating (Figure 7(a)). Tsong Yueh Chen, apart from being one of the early researchers in the ART study area, he also has the highest number of research contribution to the area as he is a co-author in 45 papers. He is followed by Fei-Ching Kuo, who has also co-authored in 28 papers in the area of study. Authors like Huai Liu, Dave Towey, Kwok Ping Chan, and Johannes Mayer, have also made very huge contributions to the area. Further investigations on the background of all the authors with very high contributions to ART revealed that most of them are co-authors to several papers in the area. Our investigation further revealed that most of these authors have a relation in some way. For example, Fei-Ching Kuo, De Hao Huang, Huai Liu, and Robert G. Merkel, had their PhD in the Swinburne University of Technology under Tsong Yueh Chen; while Jinfu Chen, Rubing Huang, and Xiaodong Xie, had their PhD in Huazhong University of Science & Technology under Yansheng Lu. There have also been a number of collaborations among most of these authors resulting in their high contributions in the area. This is evident from the diverse locations of co-authors observed in most of the studies considered for this review. These observations suggest that the common affiliated institutions of most of the authors and their high collaboration in the study area are factors that contributed to the high numbers of ART studies among them. This provides compelling evidence that author affiliations and collaborations influence research contributions in ART studies. The research contributions to ART have been quite significant since this is shown by the high number of authors and author contributions in the area of study.

In the evaluation of any testing approach, the use of real-life programs is valuable in determining the quality of the testing approach. The analysis of the trend on the usage of subject programs shows that not much (38.5%) empirical studies with real-life programs have been done in the ART literature. In addition, only a few subject programs (12 programs) have been utilized in a high number of ART studies. Our further analysis of the studies provides evidence to suggest that the extensive use of 12 programs is because they were the first set of subject programs used to empirically study the ART technique [27]. Although several programs have been utilized in the empirical study of ART, the frequencies of their use are not adequate, as majority have been used once. We think that empirical studies with subject programs must be a requirement for all future research works on ART, in order to provide an effective validation of the capability of ART approaches.

2) RQ2–VARIATIONS OF ART METHODS AND THEIR CHARACTERISTICS

Various ART methods have been proposed in an effort to improve the failure-detection effectiveness and to reduce the overheads of the ART technique. These ART methods have different levels of performances depending on certain factors like the cost of randomly generating an input, the cost of generating or selecting an input as a test case for that particular ART algorithm, the program execution time, the failure rate, and the failure patterns of the program under test. To taxonomically evaluate the proposed ART methods based on the contents of the literature identified, we classified the various ART methods into distance-based, exclusion-based, partition-based, and alternative-based methods (see Figure 7). The choice of categories for the ART method classification was very challenging, as there were other identified and suggested categories such as *hybrid-based* methods [146, 147], *mirror-based* methods [148, 172, 173], *lattice-based* ART methods [176, 177], *search-based* ART methods [40] [150] and *test profile-based* methods, [158, 162, 164]. However, most of the ART methods identified as *hybrid-based* were also included in either the Exclusion-based [146] or the Partition-based [147] categories. We included the ART methods identified as *mirror-based* in the Partition-based category since their approaches divide the input domain into multiple partitions; while we included the ones identified as *lattice-based* and *search-based* in the Alternative-based category, as they employed strategies that are different from the distance, exclusion, or partition approaches. Lastly, for ART methods identified as *test profile*-based, we realized that they all integrated test profile strategies with some existing ART algorithms like the distance, exclusion, or partition approaches. Therefore they provided variations to the existing ART methods. Hence, we did not consider *hybrid-based*, *mirror-based*, *lattice-based*, *search*-based, and test *profile-based* as categories in the ART method classification.

We have identified and discussed the detailed characteristics of the ART methods under each method classification in Section IV-B of this study. The distance-based ART approach has seen a lot of diverse variations over the years. Although few proposed distance-based ART methods try to reduce the number of distance computation requirement in FSCS-ART, the majority of them either provide a further investigation into the distribution of test cases and the failure-detection capability of their algorithms [42, 69, 153]. Other distance-based ART methods apply their algorithms to different scenarios such as high dimensional input domain[156] and model-based testing [107].

Exclusion-based ART is one of the first approaches to ART with the proposition of RRT in 2002 which experienced further studies in 2006 [34]. Methods proposed using exclusion-based approach normally provides enhancements to the original restricted random testing (RRT) algorithm. They try to reduce the number of distance computations in each test case selection process to increase efficiency; by combining some strategies like normalization [161], filtering [147], probability [163], and forgetting [149], in order to mainly reduce the high overheads associated with the generation of acceptable test cases in RRT. Other methods in this category try to minimize the rigidity of RRT in discarding all test cases within exclusion regions, with the argument that some of them may detect software failures. To solve the rigidity problem of RRT, most of the proposed Exclusion-based ART methods assign priorities to test cases based on their position with respect to the restricted regions to increase effectiveness.

It is evident that the partition-based ART approach gained interest from the year 2004, with the proposition of the first Partition-based ART method, named ART by Dynamic Partitioning. In addition to the fundamental aim of providing an improvement in the failure-detection effectiveness of RT, most of the partition-based methods proposed try to either improve on failure-detection effectiveness or minimize the computational overhead of other ART algorithms. Methods in the Partition-based ART category usually apply strategies to minimize distance computation to nearby executed test cases instead of involving all executed test cases. Alternatively, they apply strategies to choose test cases only from the restricted version of partitioned regions within the input domain. Although all the methods proposed under the Partition-based ART category divide the input domain in some way, they each employ different strategies to ensure the even spread of test cases.

Apart from the ART methods that employ the notions of distance, exclusion, and Partition, other ART methods have been proposed that utilize other varying algorithms or define other different diversity concepts to guide the selection of potential test cases within the input domain. Most of these other ART methods aim at improving the failure-detection effectiveness of RT. Few of them provide reduced computational cost or introduce algorithms that are effective for high dimensional input domains.

Generally, several improvements to ART methods have been proposed under each of the defined method classifications with specific motivations for each. A broad comparison of the common motivations for improvement to ART methods provide evidence to suggest that the common motivations for the improvement of ART under each classification of ART methods are mainly the (1) reduction of computational cost, (2) improvement of failure-detection effectiveness by increasing diversity of test cases, (3) reduction of edge preference in test case generation, and (4) application of ART for high dimensional input domains.

3) RQ3–TRENDS OF CONTRIBUTIONS AND DEVELOPMENT IN ART METHODS FOR NUMERIC PROGRAMS

From the papers considered in this study, almost all ART methods were proposed for numeric inputs. We identified 60 varying ART methods and classified them into four categories based on their notions: Distance-based, Exclusion-based, Partition-based, and Alternative-based ART methods. The methods that employ partition in their test generation strategies have gained more researcher contributions. The reason is that of the increased performance achieved mostly by combining the partition strategy with other strategies. Upon further analysis of the literature obtained, we found that the low performance in the Exclusion-based category of methods is attributed to the high effectiveness usually attained when exclusion strategies are combined with other strategies [38, 148, 168]. It is evident from the literature that a number of Partition-based methods combine exclusion strategies in their algorithms. This suggests that the high effectiveness achieved when exclusion strategies are combined with other strategies may be a contributing factor to the high results obtained for Partition-based ART methods and the lower results obtained for Exclusion-based ART methods, as demonstrated in Figure 9. The category of ART methods that apply other different diversity concepts, referred to as *Alternative-based* category, also obtained less number of method contributions. These results are almost consistent with the number of studies for each study category shown in Figure 6. The classification of ART methods using the notions employed in their spread of test cases is very relevant, as it presents a clear variation of the different ART methods. The results provide some evidence to suggest that partition-based ART methods have had the most contributions over the years.

The results in Figure 9 further provide evidence that the contributions to ART methods that apply distance, exclusion and other different diversity concepts have been relatively low over the years. These findings are consistent with their yearly number of studies in Figure 6.

Although the Distance notion is the commonest ART notion among researchers, as the first or original ART method [27, 67] was distance-based; not many variations have been proposed over the years. However, they serve as the basis for most ART application to other program inputs [72, 79, 105].

The contribution to ART methods with the notion of Exclusion seemed very promising especially in the year 2006, thus contributing to the highest number of ART methods in that year. However, the contribution to methods in that category reduced in subsequent years. The generally low contribution to the ART methods with the notion of "Exclusion" may probably be because of the unavailability of effective variations of the existing exclusion strategies. In addition, the number of contributions can be attributed to the high performance attained by other methods when they combine their strategies to the exclusion strategy, as discussed above.

Among all the ART method classifications, the partition-based category of ART methods has experienced a more inconsistent and higher number of yearly contributions on the average. That is because; multiple numbers of new partition-based ART methods are introduced in almost every year since 2004. The inconsistency in the contributions of the Partition-based ART methods may be linked to the highly diverse ways in applying partition strategies; since the majority of the identified Partition-based methods utilize different other additional strategies in their algorithms. Thus the varying number of Partition-based ART methods in almost every year.

The results for the trend in the number of ART method contributions shown in Figure 11 presents a dissimilar trend as compared to the yearly contribution of ART studies shown in Figure 5. The dissimilarity in magnitudes is as a result of (1) the observation that, not all ART studies gathered propose an ART method and, (2) the observation that some ART studies such as [164] proposed more than one ART method. The number of yearly ART methods developed has been quite constant with a record maximum recorded in the year 2006, except for the year 2014. The year 2014 just appears to be an anomalous less productive year for ART methods. The recent rise in the yearly number of methods, further suggests that more varying and improved ART strategies are yet to be developed as ideally, every ART strategy outperforms RT in terms of failure-detection capability.

**B. OPEN ISSUES IN ART**
This subsection introduces some open issues related to ART and discusses them.

1) COMPARISON OF ART METHODS TO OTHER TECHNIQUES
There is no challenge that RT is more cost-effective than any testing method under every specific testing scenario. ART methods are brought mainly to enhance RT methods in terms of effectiveness in detecting failure, and hence the focus of ART researchers shall be on the improvement of their methods over RT. Since ART is proposed as an extension to RT, it is realistic to validate an improvement by comparing any newly proposed ART method using RT method as a benchmark. In addition, ART methods present higher computational overheads than RT due to their additional mechanisms to ensure an even spread or diversity of test cases. Therefore, the focus of researchers is to propose an ART method that can provide higher failure detection effectiveness than RT, and can compare to existing ART methods in terms of effectiveness and computational overhead. Hence, the comparison made between any new ART method and RT, and sometimes other ART methods.

However, as is evident in Section IV-B of this study, varying ART methods perform differently under different scenarios. Therefore, a general comparison of ART to other testing techniques other than RT requires the choice of a specific ART method to represent the entire family of ART methods. This is because the performance of the chosen ART method will directly reflect the performance of the family of ART methods. However, some researchers have compared their proposed test generation techniques to ART under different scenarios, such as Hemmati *et al.* [182] for model-based testing and Nie *et al.* [100] for detecting interaction trigger failures. A study by Iqbal *et al.* [183] observed that ART performed best when they compared RT, ART and search-based testing using Genetic Algorithms and the (1 + 1) Evolutionary Algorithm. Their study consisted of a real-life real-time embedded seismic system. These comparisons have always considered the basic or original ART method (FSCS-ART) [27] to represent the family of ART methods, as researchers may argue that such comparisons require the use of a basic ART method. Wu *et al.* [109] have recently performed an empirical comparison of ART to RT and *Combinatorial testing* (CT) —ART outperformed RT in fault detection ability and compared to CT in 96% of test scenarios.

2) CHOICE OF AN ART METHOD FOR A TEST
The choice of a specific ART method over other ART methods is quite intricate. Whether a particular ART method is better than another depends on (at least), the cost of randomly generating a test case, the cost of selecting a test case as an input, the execution time, the failure rate, and failure pattern of the program under test. The first three items may be estimated or averaged. But, the last two items would be very difficult to be known prior to testing. It is not easy to have a good estimation of failure rate, and even more difficult to know or estimate the failure patterns. Generally, the location information and the shape of failure patterns can facilitate the test data selection process of black-box testing methods [36]. This justifies why each ART algorithms perform differently under different scenarios. Therefore, the choice of an ART method depends on the scenario under which it will be used, such as the type of program, specific failure rates, and the test generation time. We believe this dependence can be resolved with an ART method with relatively higher performances for all scenarios in the future.

3) APPLICATION OF ART IN TESTING TOOLS
The application of a testing technique in tools is an important enabler for its transformation from the laboratory into a practical and widely used testing technique. Since the first proposal of the idea of ART, not much development work on testing tools that apply the ART technique has been done. There are currently a few testing tools built to fully support automated ART analysis. For example, the ART algorithm has been fully implemented in the smart-monkey [31] tool within MobileTest, a testing framework for automatic black-box testing of mobile applications [184]. The smart-monkey tools can test the set-top box system for the play station for video games, interactive TV and sensor-driven automaton arms on condition that the input events are defined. ARTOO [28] has also been as a plug-in strategy for input generation. in the AutoTest tool [29] based on the object distance. AutoTest is a tool that performs fully automatic unit testing of Eiffel code. With such a feature, AutoTest could easily support other ART algorithms. [74] have developed the tool ARTGen that supports the testing of Java programs using a divergence-oriented approach to ART. Iqbal *et al.* [182] have developed an automated test framework which can support ART to test real-time embedded systems, and the framework has been found effective.

As we explained above, most random-based techniques like ART are integrated into other testing tools or techniques to generate test suites, especially for complex programs. An example of such integration is the adoption of ART in a structural test data generation tool known as AUgmented Search–based TestINg (AUSTIN) tool [30]. AUSTIN is a publicly available SBST tool for the C language, which is very effective and efficient. Some researches such as [7, 30, 108, 185, 186] have proven that ART can be integrated into testing tools to generate more quality and high coverage test oracles. In fact, ART's processes of generating random and even test cases across the input domain are quite simple. Hence, it is very easy to build an ART tool on top of a random test case generator. It is also quite simple to plugin the random and even spreading component of ART into an existing RT tool.

3) QUALITY TEST CASES
Arcuri and Briand [32] compared ART and RT in the generation of automated oracles. The authors concluded that when you have an automated oracle, the number of the test cases you sample is simply irrelevant: the time required to execute the test cases is the only important metric in this case. However, they never mention how the diversity of the test case sample can enhance performance especially in situations where test executions are expensive. We argue that the number of test cases sampled is also very relevant, as the time required in executing the test cases can be minimized by the speed of the test automation system. In addition, the effectiveness of detecting failure is more important than the time it takes to detect failure since computation is typically cheaper than human effort. Several empirical analyses [28, 61, 73, 165] show that ART can sample quality set of test cases (requires fewer test cases to detect failure) than RT. The work of Iqbal *et al.* [182], reported that ART yielded the best performance (100%) in terms of failure detection and high-quality test cases. In another dimension, ART studies such as the works of Barus *et al.* [66], Shahbazi *et al.* [61] and Huang *et al.* [173] have demonstrated feasible and computationally efficient schemes of linear order (similar to that of RT) for applying ART, which addresses the issue of cost-effectiveness argued by Arcuri and Briand [32].

C. FUTURE DIRECTIONS
ART methods are still undergoing several evolutions, and there remains a lot of work to be done to further explore their potential. As research is ongoing, different strategies to ART will be realized, that will have significant influences on software testing studies.

Early studies on ART have concentrated principally on numeric input domains, but recent studies have shown that it is applicable to a broad range of software. The success of ART is an indication of the potential of failure-based testing approaches, and it provides evidence of the significance of diversity in terms of influencing the effectiveness of test suites. As such, we believe that many future software testing applications that will require the generation of random test cases will consider ART, as it represents an effective, efficient alternative to RT.

The ability to easily obtain a good estimate of the failure rate and the failure pattern type of the program under test prior to testing can enhance ART approaches. Thus, more research may need to be performed to explore possible ART testing approaches based on failure. The consideration of failure in developing future ART approaches can further provide new ways to debug and to repair programs, not just restricted to testing.

The review focused on ART literature and analyzed ART methods for numeric programs. There remain some areas in this review, which requires further study. We think further research is required to separately analyze ART methods based on their specific and appropriate application areas. It will also be of essence to investigate further other metrics for the application of ART to object-oriented software, which will minimize the limitations of existing ones. In addition, it will be of extreme interest to include other researches that reside at the borderline between ART and other testing methods that select random test cases or diversify test cases in a future survey.

### D. LIMITATIONS OF THIS STUDY

Access to appropriate studies and gaps in literature are difficulties with reviews. We made efforts to follow the procedures outlined in Kitchenham's guidelines [62], though with little variance.

The initial electronic search using search strings was organized as an automated search process. This was in line with the practices of other researchers looking at research development trends. The search performed found all papers related to ART. However, since the search process was done by a single author, there is a probability that we have missed some studies; especially those that are on the borderline between ART and other testing techniques that implement even spreading of test cases.

Two authors examined every issue of each of the studies in question according to the set selection criteria; although the studies included and excluded in this review were further checked by a third author using their titles, abstracts, conclusions, and sometimes checking the content of the papers, to ensure quality and relevance. We may have probably missed articles on specific ART topics which may have possibly addressed some of the research questions.

All selected candidate studies were shared among all the authors to review and extract data from them. The extracted data were then checked and discussed among the authors and relevant data chosen for inclusion in this review. It is probable that some of the data that we collected may be erroneous as the data extraction process can lead to problems especially for complex data [187].

However, in this study, the use of the guidelines of conducting literature reviews [62] in developing and reviewing research protocol and the use of card sorting technique [68] in data elicitation improved the quality of data classification and extraction process. The research questions were used to guide the data extraction process to achieve consistent extraction of relevant information. The data extracted from the selected articles were quite objective; as such we expect very few errors in the data extraction process. Also, the constant and independent evaluation of the quality criteria by two authors hopefully reduced the possibility of erroneous results. This provides assurance that this work can be replicated by other researchers to obtain the same results.

## VI. CONCLUSION

In this paper, we have presented a comprehensive review of ART by investigating five main approaches. We have explored the trends of various literature on ART by classifying them into five categories and evaluating their study contributions. We have also reviewed ART methods proposed in the literature for numeric programs and categorized them based on the notions they employ to achieve even spread of test inputs: distance-based, exclusion-based, partition-based, and alternative-based categories. Additionally, we have analyzed the trends in the developments of ART methods to date. Finally, we discussed several worthy avenues for future investigation. This section provides some conclusions we have drawn from this review.

The evaluation of the 109 ART-related papers we identified for this review shows that studies in ART have been quite unstable but with consistent contributions from 2001 to 2018. Our investigations show that significant research efforts (49.5% of the total ART studies identified) have been made towards theoretical issues about ART such as hypotheses supporting ART, evaluation of features of ART, criticisms of ART, the proposition of distance metrics for ART, ART-related theses and reviews. This is, therefore, an indication that the theory of ART has been significantly researched. ART studies that employ the notion of *partition* have been studied more than other notions. The number of studies on ART has been inconsistent, which shows varying levels of researcher interests. The inconsistent number of ART studies for different periods provides reasons to believe that the number of software engineering related research conferences and workshops may have an influence on the contributions over the years. The stability of the recent yearly number of ART studies and the notions of the current studies suggest that more investigations on ART must be expected especially in the area of adopting the ART algorithms for diverse scenarios or as part of other testing methods.

The researcher contributions to ART studies have also been relatively significant, with 125 authors or co-authors and majority of them having one paper to their name. Tsong Yueh Chen has the highest number of research contribution to the area and has co-authored about 45 ART studies. Our investigations provide compelling evidence that author affiliation institutions and collaborations have been a contributing factor to the high numbers of researches in ART.

The number of studies that perform empirical evaluations of ART with real-life programs is quite limited, consisting of only 38.5%. Most of the studies were based on simulation, therefore future empirical studies on ART must require the use of real programs as subjects to encourage a more effective validation and their enhance generalization.

We identified 60 proposed ART methods for numeric programs, which can be categorized by certain notions they employ. Among the various categories of ART methods identified in this review, our evaluation showed that researchers have proposed more partition-based ART methods than for any other category of ART variants. In relation to the aforementioned trends in ART studies, we observed that the consistently high number of ART methods proposed each year further suggests that more varying and improved ART strategies are yet to be developed.

From the review of the various literature on ART, it is evident that ART yields the best performance when compared with RT in terms of failure detection effectiveness. In addition, the implementation of any method in the family of ART methods requires consideration of the test scenario and requirements.

Generally, the study provides evidence that the field of ART is not yet matured, although it has a relatively large number of studies on its theory and varying methods; but rather one that is devising different strategies to make ART more cost-effective and applicable in different test scenarios in order to impact on the industry. Although this review may be constrained by the data extraction process, we are confident that our careful extraction and aggregation of the data provided an overview of all the related papers in the area of study especially related to numeric programs. Our review may be used as a reference for further studies in software testing, especially for ART studies and can significantly expand on the knowledge of software engineering.

## ACKNOWLEDGMENT

We would like to thank T. Y. Chen for his helpful comments in an earlier version of this article.

## REFERENCES

[1] G. J. Myers, *Software Reliability*, New York: John Wiley & Sons, Inc., 1976.

[2] B. Hailpern and P. Santhanam, "Software debugging, testing, and verification," *IBM Systems Journal,* vol. 41, no. 1, pp. 4-12, 2002. doi:10.1147/sj.411.0004.

[3] R. Blanco, J. Tuya, and B. Adenso-Díaz, "Automated test data generation using a scatter search approach," *Information and Software Technology,* vol. 51, no. 4, pp. 708-720, 2009. doi:10.1016/j.infsof.2008.11.001.

[4] R. Hamlet, "Random testing," *Encyclopedia of Software Engineering*, J. Marciniak, ed., pp. 970–978, New York, NY: John Wiley & Sons, 2002. doi:10.1002/0471028959.sof268.

[5] T. Y. Chen, S. C. Cheung, and S. M. Yiu, *Metamorphic testing: a new approach for generating next test cases*, Department of Computer Science, Hong Kong University of Science and Technology, Hong Kong, 1998.

[6] P. McMinn, "Search-based software test data generation: a survey," *Software testing, Verification and reliability,* vol. 14, no. 2, pp. 105-156, 2004. doi:10.1002/stvr.294.

[7] M. Z. Iqbal, A. Arcuri, and L. Briand, "Combining search-based and adaptive random testing strategies for environment model-based testing of real-time embedded systems," in *Proceedings of the International Symposium on Search Based Software Engineering (SSBSE, 2012)*, Riva del Garda, Trento, Italy, 2012*,* pp. 136-151. doi:10.1007/978-3-642-33119-0_11.

[8] D. Hamlet and R. Taylor, "Partition testing does not inspire confidence (program testing)," *IEEE Transactions on Software Engineering,* vol. 16, no. 12, pp. 1402-1411, 1990. doi:10.1109/32.62448.

[9] R. Kuhn, R. Kacker, Y. Lei, and J. Hunter, "Combinatorial software testing," *Computer,* vol. 42, no. 8, 2009.

[10] O. Bühler and J. Wegener, "Evolutionary functional testing," *Computers & Operations Research,* vol. 35, no. 10, pp. 3144-3160, 2008. doi:10.1016/j.cor.2007.01.015.

[11] L. J. White, "Software testing and verification," *Advances in computers,* vol. 26, pp. 335-391, 1987. doi:10.1016/S0065-2458(08)60010-8.

[12] T. Yoshikawa, K. Shimura, and T. Ozawa, "Random program generator for Java JIT compiler test system," in *Proceedings of the Third International Conference on Quality Software (QSIC 2003)*, 2003, pp. 20-23. doi:10.1109/QSIC.2003.1319081.

[13] B. P. Miller, D. Koski, C. P. Lee, V. Maganty, R. Murthy, A. Natarajan, and J. Steidl, *Fuzz revisited: A re-examination of the reliability of UNIX utilities and services*, Technical report, 1995.

[14] K. Sen, "Race directed random testing of concurrent programs," *ACM SIGPLAN Notices. ACM,* vol. 43, no. 6, pp. 11-21, 2008. doi:10.1145/1379022.1375584.

[15] D. Owen, T. Menzies, M. Heimdahl, and J. Gao, "Finding faults quickly in formal models using random search." Presented at the *SEKE* 2003.

[16] J. E. Forrester and B. P. Miller, "An empirical study of the robustness of Windows NT applications using random testing," in *Proceedings of the 4th USENIX Windows System Symposium*, Seattle, Washington, August, 2000*,* pp. 59-68.

[17] D. L. Bird and C. U. Munoz, "Automatic generation of random self-checking test cases," *IBM systems journal,* vol. 22, no. 3, pp. 229-245, 1983. doi:10.1147/sj.223.0229.

[18] P. Godefroid, N. Klarlund, and K. Sen, "DART: directed automated random testing," in *Proceedings of the 2005 ACM SIGPLAN conference on Programming language design and implementation*, Chicago, IL, USA, June, 2005*,* pp. 213-223. doi:10.1145/1064978.1065036.

[19] G. J. Myers, *The Art of Software Testing. Revised and updated by T. Badgett, T.M. Thomas and C. Sandler, Hoboken*: NJ: John Wiley and Sons, 2004.

[20] J. W. Duran and S. C. Ntafos, "An evaluation of random testing," *IEEE transactions on Software Engineering,* vol. SE-10, no. 4, pp. 438-444, 1984. doi:10.1109/TSE.1984.5010257.

[21] P. G. Bishop, "The Variation of Software Survival Time for Different Operational Input Profiles (or why you can wait a long time for a big bug to fail)." Presented at the *Twenty-Third International Symposium on Fault-Tolerant Computing (FTCS-23 Digest of Papers, 1993).* Toulouse, France, 1993, pp. 98-107. doi:10.1109/FTCS.1993.627312.

[22] P. E. Ammann and J. C. Knight, "Data diversity: An approach to software fault tolerance," *IEEE Transactions on Computers,* vol. 37, no. 4, pp. 418-425, 1988. doi:10.1109/12.2185.

[23] F. T. Chan, T. Y. Chen, I. K. Mak, and Y.-T. Yu, "Proportional sampling strategy: guidelines for software testing practitioners," *Information and Software Technology,* vol. 38, no. 12, pp. 775-782, 1996. doi:10.1016/0950-5849(96)01103-2.

[24] L. J. White and E. I. Cohen, "A domain strategy for computer program testing," *IEEE Transactions on Software Engineering,* vol. 6, no. 3, pp. 247-257, 1980. doi:10.1109/TSE.1980.234486.

[25] G. B. Finelli, "NASA software failure characterization experiments," *Reliability Engineering & System Safety,* vol. 32, no. 1, pp. 155-169, 1991. doi:10.1016/0951-8320(91)90052-9.

[26] C. Schneckenburger and J. Mayer, "Towards the determination of typical failure patterns." Presented at the *Fourth international workshop on Software quality assurance (SOQUA'07) in conjunction with the 6th ESEC/FSE joint meeting*, Dubrovnik, Croatia, 2007, pp. 90-93. doi:10.1145/1295074.1295091.

[27] T. Y. Chen, H. Leung, and I. K. Mak, "Adaptive Random Testing," in *Proceedings of the Ninth Asian Computing Science Conference (ASIAN'04), Lecture Notes in Computer Science*, Chiang Mai, Thailand, 2004, pp. 320-329. doi:10.1007/978-3-540-30502-6_23.

[28] I. Ciupa, A. Leitner, M. Oriol, and B. Meyer, "Artoo: adaptive random testing for object-oriented software," in *Proceedings of the 30th ACM/IEEE International Conference on Software Engineering*, Leipzig, Germany, May, 2008, pp. 71-80. doi:10.1109/ICST.2008.20.

[29] A. LEITNER and I. CIUPA, "AutoTest," http://se.inf.ethz.ch/people/leitner/auto test/, 2005 - 2007.

[30] K. Lakhotia, M. Harman, and H. Gross, "AUSTIN: A tool for search based software testing for the C language and its evaluation on deployed automotive systems," in *Proceedings of the Search Based Software Engineering (SSBSE), 2010 Second International Symposium on*, 2010, pp. 101-110. doi:10.1109/SSBSE.2010.21.

[31] Z. Liu, X. Gao, and X. Long, "Adaptive random testing of mobile application," in *Proceedings of the 2nd International Conference on Computer Engineering and Technology (ICCET)*, 2010, pp. V2-297-V2-301. doi:10.1109/ICCET.2010.5485442.

[32] A. Arcuri and L. Briand, "Adaptive random testing: An illusion of effectiveness?," in *Proceedings of the 2011 International Symposium on Software Testing and Analysis*, 2011, pp. 265-275. doi:10.1145/2001420.2001452.

[33] K. P. Chan, T. Y. Chen, and D. Towey, "Restricted random testing." Presented at the *Software Quality—ECSQ 2002*, 2002, pp. 321-330. doi:DOI: 10.1007/3-540-47984-8_35.

[34] K. P. Chan, T. Y. Chen, and D. Towey, "Restricted random testing: Adaptive random testing by exclusion," *International Journal of Software Engineering and Knowledge Engineering,* vol. 16, no. 04, pp. 553-584, 2006. doi:10.1142/S0218194006002926.

[35] J. Mayer, "Restricted Adaptive Random Testing by Random Partitioning." Presented at the *Software Engineering Research and Practice*, 2006, pp. 59-65.

[36] C. Mao, "Adaptive random testing based on two-point partitioning," *International Journal of Informatica,* vol. 36, no. 3, 2012.

[37] C. Chow, T. Y. Chen, and T. H. Tse, "The art of divide and conquer: an innovative approach to improving the efficiency of adaptive random testing," in *Proceedings of the 13th International Conference on Quality Software (QSIC, 2013)*, Nanjing, China, 2013, pp. 268-275. doi:10.1109/QSIC.2013.19.

[38] K. k. sabor and m. mohsenzadeh, "Adaptive random testing through dynamic partitioning by localization with restriction and enlarged input domain," in *Proceedings of the 2012 International Conference on Information Technology and Software Engineering: Software Engineering & Digital Media Technology*, 2013, pp. 147-155. doi:10.1007/978-3-642-34531-9_16.

[39] C. Mao, T. Y. Chen, and F.-C. Kuo, "Out of sight, out of mind: a distance-aware forgetting strategy for adaptive random testing," *Science China Information Sciences,* vol. 60, no. 9, pp. 092106, 2017. doi:10.1007/s11432-016-0087-0.

[40] C. Schneckenburger and F. Schweiggert, "Investigating the dimensionality problem of Adaptive Random Testing incorporating a local search technique," in *Proceedings of the IEEE International Conference on Software Testing Verification and Validation Workshop, 2008 (ICSTW'08)* 2008, pp. 241-250. doi:10.1109/ICSTW.2008.24.

[41] T. Y. Chen, T. H. Tse, and Y.-T. Yu, "Proportional sampling strategy: a compendium and some insights," *Journal of Systems and Software,* vol. 58, no. 1, pp. 65-81, 2001. doi:10.1016/S0164-1212(01)00028-0.

[42] R. G. Merkel, "Analysis and enhancements of adaptive random testing," PhD thesis, School of Information Technology, Swinburne University of Technology, 2005.

[43] D. Towey, "Studies of different variations of adaptive random testing," PhD Dissertation, University of Hong Kong, 2006.

[44] T. A. Thayer, M. Lipow, and E. C. Nelson, *Software Reliability-A Study of Large Project Reality (TRW series of software technology)*, North-Holland, Amsterdam, The Netherlands, 1978.

[45] M. J. P. Van der Meulen, P. G. Bishop, and R. Villa, "An exploration of software faults and failure behaviour in a large population of programs," in *Proceedings of the 15th International Symposium on Software Reliability Engineering, 2004. (ISSRE 2004)*, 2004, pp. 101-112. doi:10.1109/ISSRE.2004.7.

[46] A. M. Sinaga, "Applying Feedback Information for Random Partition Testing," in *Proceedings of the 2018 IEEE/ACIS 17th International Conference on Computer and Information Science (ICIS)*, Singapore, 6-8 Jun., 2018, pp. 224-228. doi:10.1109/ICIS.2018.8466515.

[47] I. K. Mak, "On the effectiveness of random testing," Doctoral dissertation, University of Melbourne, Faculty of Science, 1998.

[48] J. Mayer, "Adaptive random testing by bisection and localization," in *Proceedings of the International Workshop on Formal Approaches to Software Testing*, 2005, pp. 72-86. doi:10.1007/11759744_6.

[49] T. Y. Chen, D. H. Huang, and Z. Q. Zhou, "Adaptive random testing through iterative partitioning," in *Proceedings of the 11th Ada-Europe International Conference on Reliable Software Technologies – Ada-Europe 2006*, Porto, Portugal, June, 2006, pp. 155-166. doi:10.1007/11767077_1.

[50] Y. K. Malaiya, "Antirandom testing: Getting the most out of black-box testing," in *Proceedings of the Sixth International Symposium on Software Reliability Engineering, 1995*, 1995, pp. 86-95. doi:10.1109/ISSRE.1995.497647.

[51] T. Y. Chen and R. G. Merkel, "Quasi-random testing," *IEEE Transactions on Reliability,* vol. 56, no. 3, pp. 562-568, 2007. doi:10.1109/TR.2007.903293.

[52] H. Liu and T. Y. Chen, "An innovative approach to randomising quasi-random sequences and its application into software testing," in *Proceedings of the Quality Software, 2009. QSIC'09. 9th International Conference on*, Jeju, South Korea, Aug. 24-25 2009, pp. 59-64. doi:10.1109/QSIC.2009.16.

[53] H. Leung, T. H. Tse, F. Chan, and T. Y. Chen, "Test case selection with and without replacement," *Information Sciences,* vol. 129, no. 1-4, pp. 81-103, 2000. doi:10.1016/S0020-0255(00)00059-1.

[54] E. J. Weyuker and B. Jeng, "Analyzing partition testing strategies," *IEEE transactions on software engineering,* vol. 17, no. 7, pp. 703-711, 1991. doi:10.1109/32.83906.

[55] T. Y. Chen and R. G. Merkel, "An upper bound on software testing effectiveness," *ACM Transactions on Software Engineering and Methodology (TOSEM),* vol. 17, no. 3, pp. 16, 2008. doi:10.1145/1363102.1363107.

[56] T. Y. Chen and Y.-T. Yu, "On the relationship between partition and random testing," *IEEE Transactions on Software Engineering,* vol. 20, no. 12, pp. 977-980, 1994. doi:10.1109/32.368132.

[57] T. Y. Chen, F.-C. Kuo, and R. G. Merkel, "On the statistical properties of the f-measure," in *Proceedings of the Fourth International Conference on Quality Software (QSIC 2004)*, Braunschweig, Germany, 2004, pp. 146-153. doi:10.1109/QSIC.2004.1357955.

[58] F.-C. Kuo, T. Y. Chen, H. Liu, and W. K. Chan, "Enhancing adaptive random testing in high dimensional input domains," in *Proceedings of the 2007 ACM symposium on Applied computing*, 2007, pp. 1467-1472. doi:10.1145/1244002.1244316.

[59] T. Y. Chen, F.-C. Kuo, H. Liu, and W. E. Wong, "Code coverage of adaptive random testing," *IEEE Transactions on Reliability,* vol. 62, no. 1, pp. 226-237, 2013. doi:10.1109/TR.2013.2240898.

[60] T. Y. Chen, F.-C. Kuo, and Z. Q. Zhou, "On favourable conditions for adaptive random testing," *International Journal of Software Engineering and Knowledge Engineering,* vol. 17, no. 06, pp. 805-825, 2007. doi:10.1142/S0218194007003501.

[61] A. Shahbazi, A. F. Tappenden, and J. Miller, "Centroidal voronoi tessellations-a new approach to random testing," *IEEE Transactions on Software Engineering,* vol. 39, no. 2, pp. 163-183, 2013. doi:10.1109/TSE.2012.18.

[62] B. Kitchenham, *Guidelines for performing Systematic Literature Reviews in Software Engineering*, Keele University and University of Durham, 2007.

[63] J. Webster and R. T. Watson, "Analyzing the past to prepare for the future: Writing a literature review," *MIS quarterly*, pp. xiii-xxiii, 2002.

[64] J. Chen, L. Zhu, T. Y. Chen, D. Towey, F.-C. Kuo, R. Huang, and Y. Guo, "Test case prioritization for object-oriented software: An adaptive random sequence approach based on clustering," *Journal of Systems and Software,* vol. 135, pp. 107-125, 2018. doi:10.1016/j.jss.2017.09.031.

[65] X. Zhang, X. Xie, and T. Y. Chen, "Test Case Prioritization Using Adaptive Random Sequence with Category-Partition-Based Distance." Presented at the *2016 IEEE International Conference on Software Quality, Reliability and Security (QRS),* 2016, pp. 374-385. doi:10.1109/QRS.2016.49.

[66] A. C. Barus, T. Y. Chen, F.-C. Kuo, H. Liu, R. G. Merkel, and G. Rothermel, "A cost-effective random testing method for programs with non-numeric inputs," *IEEE Transactions on Computers,* vol. 65, no. 12, pp. 3509-3523, 2016. doi:10.1109/TC.2016.2547380.

[67] F. T. Chan, K. P. Chan, T. Y. Chen, and S.-M. Yiu, "Adaptive random testing with cg constraint," in *Proceedings of the 28th Annual International Conference on Computer Software and Applications (COMPSAC 2004)* Hong Kong, China, 2004, pp. 96-99. doi:10.1109/CMPSAC.2004.1342685.

[68] N. Nurmuliani, D. Zowghi, and S. P. Williams, "Using card sorting technique to classify requirements change." Presented at the *12th IEEE International Conference on Requirements Engineering* 2004, pp. 240-248. doi:10.1109/ICRE.2004.1335681.

[69] M. Patrick and Y. Jia, "Kernel Density Adaptive Random Testing," in *Proceedings of the 2015 IEEE Eighth International Conference on Software Testing, Verification and Validation Workshops (ICSTW)*, 2015, pp. 1-10. doi:10.1109/ICSTW.2015.7107451.

[70] M. Patrick and Y. Jia, "KD-ART: Should we intensify or diversify tests to kill mutants?," *Information and Software Technology,* vol. 81, no. 36-51, 2016. doi:10.1016/j.infsof.2016.04.009.

[71] A. M. Sinaga, "Adaptive Random Testing with Coverage Information for Object Oriented Program," *Advanced Science Letters,* vol. 23, no. 5, pp. 4359-4362, 2017. doi:10.1166/asl.2017.8338.

[72] J. Chen, F.-C. Kuo, T. Y. Chen, D. Towey, C. Su, and R. Huang, "A Similarity Metric for the Inputs of OO Programs and Its Application in Adaptive Random Testing," *IEEE Transactions on Reliability,* vol. 66, no. 2, pp. 373-402, 2017. doi:10.1109/TR.2016.2628759.

[73] I. P. E. S. Putra and P. Mursanto, "Centroid Based Adaptive Random Testing for object-oriented program," in *Proceedings of the 2013 International Conference on Advanced Computer Science and Information Systems (ICACSIS)*, 2013, pp. 39-45. doi:10.1109/ICACSIS.2013.6761550.

[74] Y. Lin, X. Tang, Y. Chen, and J. Zhao, "A divergence-oriented approach to adaptive random testing of Java programs," in *Proceedings of the Proceedings of the 2009 IEEE/ACM International Conference on Automated Software Engineering*, 2009, pp. 221-232. doi:10.1109/ASE.2009.13.

[75] Z. Q. Zhou, "Using coverage information to guide test case selection in adaptive random testing." Presented at the *34th IEEE Annual Computer Software and Applications Conference Workshops (COMPSACW, 2010)*, 2010, pp. 208-213. doi:10.1109/COMPSACW.2010.43.

[76] T. Y. Chen, F.-C. Kuo, R. G. Merkel, and T. H. Tse, "Adaptive random testing: The art of test case diversity," *Journal of Systems and Software,* vol. 83, no. 1, pp. 60-66, 2010. doi:10.1016/j.jss.2009.02.022.

[77] T. Y. Chen, F.-C. Kuo, and Z. Zhou, "On the Relationships between the Distribution of Failure-Causing Inputs and Effectiveness of Adaptive Random Testing," in *Proceedings of the 17th International Conference on Software Engineering and Knowledge Engineering (SEKE)*, Taipei, Taiwan, China, 2005, pp. 306-311.

[78] B. Jiang, Z. Zhang, W. K. Chan, and T. H. Tse, "Adaptive random test case prioritization," in *Proceedings of the 24th IEEE/ACM International Conference on Automated Software Engineering (ASE'09)*, Auckland, New Zealand, November, 2009, pp. 233-244. doi:10.1109/ASE.2009.77.

[79] F.-C. Kuo, "On adaptive random testing," PhD Dissertation, Swinburne University of Technology, Faculty of Information & Communication Technologies, 2006.

[80] J. Mayer and C. Schneckenburger, "Adaptive random testing with enlarged input domain," in *Proceedings of the Sixth International Conference on Quality Software, 2006. (QSIC 2006)* 2006, pp. 251-258. doi:10.1109/QSIC.2006.8.

[81] J. Mayer, "Adaptive random testing with randomly translated failure region," in *Proceedings of the 1st international workshop on Random testing*, 2006, pp. 70-77. doi:10.1145/1145735.1145746.

[82] J. Mayer and C. Schneckenburger, "An empirical analysis and comparison of random testing techniques," in *Proceedings of the Proceedings of the 2006 ACM/IEEE international symposium on Empirical software engineering*, 2006, pp. 105-114. doi:10.1145/1159733.1159751.

[83] T. Y. Chen and F.-C. Kuo, "Is adaptive random testing really better than random testing?," in *Proceedings of the 1st international workshop on Random testing*, Portland, Maine, 2006, pp. 64-69. doi:10.1145/1145735.1145745.

[84] J. Mayer, "Towards effective adaptive random testing for higher-dimensional input domains." Presented at the *8th annual conference on Genetic and evolutionary computation*, 2006, pp. 1955-1956. doi:10.1145/1143997.1144323.

[85] D. Towey, "Adaptive Random Testing: Ubiquitous Testing to Support Ubiquitous Computing," *Conference of Korea Information Technology Application Society*, pp. 138-138, 2007.

[86] K. P. Chan, T. Y. Chen, and D. Towey, "Controlling Restricted Random Testing: An Examination of the Exclusion Ratio Parameter," in *Proceedings of the Proceedings of the 19th International Conference on Software Engineering and Knowledge Engineering (SEKE 2007)*, 2007.

[87] F.-C. Kuo, K. Y. Sim, C.-a. Sun, S.-F. Tang, and Z. Zhou, "Enhanced random testing for programs with high dimensional input domains," 2007.

[88] T. Y. Chen, F.-C. Kuo, and H. Liu, "On Test Case Distributions of Adaptive Random Testing." Presented at the *19th International Conference on Software Engineering and Knowledge Engineering*, Boston, United States July 2007, pp. 141-144.

[89] Y. Liu and H. Zhu, "An experimental evaluation of the reliability of adaptive random testing methods," in *Proceedings of the Second International Conference on Secure System Integration and Reliability Improvement, 2008 (SSIRI'08)*, 2008, pp. 24-31. doi:10.1109/SSIRI.2008.18.

[90] T. Y. Chen, F.-C. Kuo, H. Liu, and W. E. Wong, "Does adaptive random testing deliver a higher confidence than random testing?," in *Proceedings of the Eighth International Conference on Quality Software, 2008. (QSIC'08)* 2008, pp. 145-154. doi:10.1109/QSIC.2008.23.

[91] H. Liu, "On Even Spread of Test Cases in Adaptive Random Testing," Swinburne University of Technology, Faculty of Information & Communication Technologies, Citeseer, 2008.

[92] A. C. Barus, "An In-depth Study of Adaptive Random Testing for Testing Program with Complex Input Types," Ph. D. dissertation, Swinburne University of Technology, 2010.

[93] Y. Yuan, Z. Fanping, Z. Guanmiao, D. Chaoqiang, and X. Neng, "Test case generation based on program invariant and adaptive

random algorithm," *Advances in Information Technology and Education*, pp. 274-282, 2011. doi:10.1007/978-3-642-22418-8_38.

[94] H. Liu, F. C. Kuo, and T. Y. Chen, "Comparison of adaptive random testing and random testing under various testing and debugging scenarios," *Software: Practice and Experience,* vol. 42, no. 8, pp. 1055-1074, 2012. doi:10.1002/spe.1113.

[95] Z. Q. Zhou, A. Sinaga, and W. Susilo, "On the fault-detection capabilities of adaptive random test case prioritization: Case studies with large test suites," in *Proceedings of the 45th Hawaii International Conference on System Science, 2012. (HICSS)* 2012, pp. 5584-5593. doi:10.1109/HICSS.2012.454.

[96] R. Huang, X. Xie, J. Chen, and Y. Lu, "Failure-detection capability analysis of implementing parallelism in adaptive random testing algorithms," in *Proceedings of the 28th Annual ACM Symposium on Applied Computing*, Coimbra, Portugal March, 2013, pp. 1049-1054. doi:10.1145/2480362.2480562.

[97] R. Huang, J. Chen, Z. Li, R. Wang, and Y. Lu, "Adaptive random prioritization for interaction test suites," in *Proceedings of the 29th Annual ACM Symposium on Applied Computing*, Gyeongju, Republic of Korea, March, 2014, pp. 1058-1063. doi:10.1145/2554850.2554854.

[98] E. Selay, Z. Q. Zhou, and J. Zou, "Adaptive random testing for image comparison in regression web testing," in *Proceedings of the 2014 International Conference on Digital lmage Computing: Techniques and Applications (DlCTA)*, 2014, pp. 1-7. doi:10.1109/DICTA.2014.7008093.

[99] T. Y. Chen, F.-C. Kuo, D. Towey, and Z. Zhou, "A revisit of three studies related to random testing," *Science China Information Sciences,* vol. 58, no. 5, 2015. doi:10.1007/s11432-015-5314-x.

[100] C. Nie, H. Wu, X. Niu, F.-C. Kuo, H. Leung, and C. J. Colbourn, "Combinatorial testing, random testing, and adaptive random testing for detecting interaction triggered failures," *Information and Software Technology,* vol. 62, pp. 198-213, 2015. doi:10.1016/j.infsof.2015.02.008.

[101] H. Liu, F.-C. Kuo, and T. Y. Chen, "Dynamic Test Profiles in Adaptive Random Testing: A Case Study." Presented at the *Software Engineering and Knowledge Engineering (SEKE 2009)*, 2009, pp. 418-421.

[102] D.-S. Koo and Y. B. Park, "OFART: OpenFlow-Switch Adaptive Random Testing." Presented at the *Advanced Multimedia and Ubiquitous Engineering*, 2017, pp. 193-198. doi:10.1007/978-981-10-5041-1_33.

[103] Y. Qi, Z. Wang, and Y. Yao, "Influence of the Distance Calculation Error on the Performance of Adaptive Random Testing," in *Proceedings of the 2017 IEEE International Conference on Software Quality, Reliability and Security Companion (QRS-C)*, Prague, Czech Republic, 2017, pp. 316-319. doi:10.1109/QRS-C.2017.60.

[104] I. Ciupa, A. Leitner, M. Oriol, and B. Meyer, "Object distance and its application to adaptive random testing of object-oriented programs," in *Proceedings of the 1st international workshop on Random testing*, Portland, Maine, July, 2006, pp. 55-63. doi:10.1145/1145735.1145744.

[105] R. Huang, X. Xie, T. Y. Chen, and Y. Lu, "Adaptive random test case generation for combinatorial testing," in *Proceedings of the 36th IEEE Annual Conference on Computer Software and Applications, (COMPSAC 2012)* Izmir, Turkey, 2012, pp. 52-61. doi:10.1109/COMPSAC.2012.15.

[106] R. Huang, J. Chen, and Y. Lu, "Adaptive Random Testing with Combinatorial Input Domain," *The Scientific World Journal*, 2014. doi:10.1155/2014/843248.

[107] B. Sun, Y. Dong, and H. Ye, "On enhancing adaptive random testing for AADL model," in *Proceedings of the 9th International Conference on Ubiquitous Intelligence & Computing and 9th International Conference on Autonomic & Trusted Computing (UIC/ATC, 2012)*, 2012, pp. 455-461. doi:10.1109/UIC-ATC.2012.77.

[108] H. Jaygarl, C. K. Chang, and S. Kim, "Practical extensions of a randomized testing tool," in *Proceedings of the 33rd Annual IEEE International Computer Software and Applications Conference (COMPSAC '09)*, Seattle, WA, USA, July, 2009, pp. 148-153. doi:10.1109/COMPSAC.2009.29.

[109] H. Wu, J. Petke, Y. Jia, and M. Harman, "An Empirical Comparison of Combinatorial Testing, Random Testing and Adaptive Random Testing," *IEEE Transactions on Software Engineering*, pp. 1-1, Jul. 06 2018. doi:10.1109/TSE.2018.2852744.

[110] R. M. Sidek, A. A. A. Ghani, H. Zulzalil, S. Baharom, and A. Noraziah, "A Code Profiling Using Statistical Testing in StART," *Advanced Science Letters,* vol. 24, no. 10, pp. 7295-7299, 2018. doi:10.1166/asl.2018.12931.

[111] L. Ma, P. Wu, and T. Y. Chen, "Diversity driven adaptive test generation for concurrent data structures," *Information and Software Technology,* vol. 103, pp. 162-173, Nov. 2018. doi:10.1016/j.infsof.2018.07.001.

[112] Y. Jia and M. Harman, "An analysis and survey of the development of mutation testing," *IEEE transactions on software engineering,* vol. 37, no. 5, pp. 649-678, 2011. doi:10.1109/TSE.2010.62.

[113] B. P. Flannery, W. H. Press, S. A. Teukolsky, and W. T. Vetterling, *Numerical recipes 3rd edition: The art of scientific computing*, 3rd ed.: Cambridge university press, 2007.

[114] ACM, "Collected Algorithms from ACM: Volume 1, 2, 3," *New York: Association for Computer Machinery, 1980*.

[115] "SIR: Software-artifact Infrastructure Repository," 2018; http://sir.unl.edu/.

[116] "GNU Scientific Library. A product of the GNU project.," 2008; http://www.gnu.org/software/gsl/.

[117] "Junit, Testing Resources for Extreme Programming," 2009; http://www.junit.org/.

[118] K. Thompson and D. M. Ritchie, *unix Programmer's Manual* (Fifth Edition) ed.: Bell Telephone Laboratories, 1975.

[119] C. Lott. "A repeatable software engineering experiment," 2013; http://www.maultech.com/chrislott/work/exp/.

[120] T. Y. Chen, F.-C. Kuo, Y. Liu, and A. Tang, "Metamorphic Testing and Testing with Special Values," in *Proceedings of the SNPD*, 2004, pp. 128-134.

[121] T. H. Cormen, C. E. Leiserson, R. L. Rivest, and C. Stein, *Introduction to algorithms second edition*: The MIT Press, 2001.

[122] A. B. Sánchez, S. Segura, J. A. Parejo, and A. Ruiz-Cortés, "Variability testing in the wild: the Drupal case study," *Software & Systems Modeling,* vol. 16, no. 1, pp. 173-194, Apr. 19 2017. doi:10.1007/s10270-015-0459-z.

[123] F. Medeiros, C. Kästner, M. Ribeiro, R. Gheyi, and S. Apel, "A comparison of 10 sampling algorithms for configurable systems." Presented at the *38th International Conference on Software Engineering*, Austin, Texas, May 14-22 2016, pp. 643-654. doi:10.1145/2884781.2884793.

[124] "GitHub, where software is built," 2015; https://github.com.

[125] W. Visser, C. S. Păsăreanu, and R. Pelánek, "Test input generation for java containers using state matching," in *Proceedings of the Proceedings of the 2006 international symposium on Software testing and analysis*, 2006, pp. 37-48. doi:10.1145/1146238.1146243.

[126] "Package java.util," 2018; https://docs.oracle.com/javase/7/docs/api/java/util/package-summary.html.

[127] "Asm: Java bytecode manipulation and analysis framework," 2009; http://asm.objectweb.org/.

[128] "Apache Ant, The Apache Ant project," 2018; https://ant.apache.org/.

[129] "Codeforge-free open source codes forge and sharing," 2013; http://www.codeforge.com,.

[130] "Sourceforge-download, develop and publish free open source software," 2013; http://sourceforge.net.

[131] "Codeplex-open source project hosting," 2013; http://www.codeplex.com.

[132] "Codeproject - for those who code," 2013; http://www.codeproject.com.

[133] "Apache Commons," 2012; http://commons.apache.org/.

[134] "The EiffelBase Library. Eiffel Software Inc.," 2008; http://www.eiffel.com/.

[135] M. Herlihy and N. Shavit, *The art of multiprocessor programming*: Morgan Kaufmann, 2011.

[136] M. Herlihy, Y. Lev, V. Luchangco, and N. Shavit, "A simple optimistic skiplist algorithm," in *Proceedings of the International Colloquium on Structural Information and Communication Complexity*, 2007, pp. 124-138.

[137] S. Heller, M. Herlihy, V. Luchangco, M. Moir, W. N. Scherer, and N. Shavit, "A lazy concurrent list-based set algorithm," in *Proceedings of the International Conference On Principles Of Distributed Systems*, 2005, pp. 3-16.

[138] M. M. Michael and M. L. Scott, *Simple, fast, and practical non-blocking and blocking concurrent queue algorithms*, ROCHESTER UNIV NY DEPT OF COMPUTER SCIENCE, 1995.

[139] "A fast no lock RingQueue with multi threads," 2017; https://github.com/dodng/fast_ring_queue.

[140] D. Hendler, N. Shavit, and L. Yerushalmi, "A scalable lock-free stack algorithm," in *Proceedings of the Proceedings of the sixteenth annual ACM symposium on Parallelism in algorithms and architectures*, 2004, pp. 206-215.

[141] "Concurrent hashmap implementation with striped lock," 2017; https://github.com/kumagi/hashmap.

[142] H. Do, S. Elbaum, and G. Rothermel, "Supporting controlled experimentation with testing techniques: An infrastructure and its potential impact," *Empirical Software Engineering,* vol. 10, no. 4, pp. 405-435, 2005. doi:10.1007/s10664-005-3861-2.

[143] M. Hutchins, H. Foster, T. Goradia, and T. Ostrand, "Experiments of the effectiveness of dataflow-and controlflow-based test adequacy criteria," in *Proceedings of the Proceedings of the 16th international conference on Software engineering (ICSE '94)*, Sorrento, Italy, 1994, pp. 191-200.

[144] M. Galassi, J. Davies, J. Theiler, B. Gough, G. Jungman, P. Alken, M. Booth, and F. Rossi, "GNU scientific library," *Network Theory Ltd,* vol. 3, 2002.

[145] S.-S. Hou, C. Zhang, D. Hao, and L. Zhang, "PathART: path-sensitive adaptive random testing," in *Proceedings of the 5th Asia-Pacific Symposium on Internetware*, Changsha, China, October, 2013, p. 23. doi:10.1145/2532443.2532460.

[146] T. Y. Chen, R. G. Merkel, P.-K. Wong, and G. Eddy, "Adaptive Random Testing Through Dynamic Partitioning," in *Proceedings of the Fourth International Conference on Quality Software (QSIC 2004)* Braunschweig, Germany, 2004, pp. 79-86. doi:10.1109/QSIC.2004.1357947.

[147] K. P. Chan, T. Y. Chen, and D. Towey, "Adaptive random testing with filtering: An overhead reduction technique," in *Proceedings of the 17th International Conference on Software Engineering and Knowledge Engineering (SEKE'05)*, Taipei, Taiwan, 2005, pp. 14-16.

[148] K. P. Chan, T. Y. Chen, F.-C. Kuo, and D. Towey, "A revisit of adaptive random testing by restriction." Presented at the *28th Annual International Computer Software and Applications Conference (COMPSAC 2004)*, Hong Kong, China, September 2004, pp. 78-85. doi:10.1109/CMPSAC.2004.1342809.

[149] K. P. Chan and D. Towey, "Forgetting test cases," in *Proceedings of the 30th Annual International Computer Software and Applications Conference (COMPSAC'06)*, Chicago, USA, 2006, pp. 485-494. doi:10.1109/COMPSAC.2006.43.

[150] A. F. Tappenden and J. Miller, "A novel evolutionary approach for adaptive random testing," *IEEE Transactions on Reliability,* vol. 58, no. 4, pp. 619-633, 2009. doi:10.1109/TR.2009.2034288.

[151] K. P. Chan, T. Y. Chen, and D. Towey, "Good random testing," in *Proceedings of the International Conference on Reliable Software Technologies (Ada-Europe 2004)*, Palma de Mallorca, Spain, June, 2004, pp. 200-212. doi:10.1007/978-3-540-24841-5_16.

[152] J. Mayer and C. Schneckenburger, "Statistical Analysis and Enhancement of Random Testing Methods also under Constrained Resources," *Software Engineering Research and Practice,* pp. 16-23, 2006.

[153] T. Y. Chen, F.-C. Kuo, and H. Liu, "Adaptive random testing based on distribution metrics," *Journal of Systems and Software,* vol. 82, no. 9, pp. 1419-1433, 2009. doi:10.1016/j.jss.2009.05.017.

[154] T. Y. Chen, F.-C. Kuo, and H. Liu, "Enhancing adaptive random testing through partitioning by edge and centre," in *Proceedings of the 18th Australian Software Engineering Conference, 2007 (ASWEC 2007)* Melbourne, Vic., Australia, 2007, pp. 265-273. doi:10.1109/ASWEC.2007.20.

[155] A. M. Sinaga, O. D. Hutajulu, R. T. Hutahaean, and I. C. Hutagaol, "Path Coverage Information for Adaptive Random Testing," in *Proceedings of the 2017 International Conference on Information Technology*, Singapore, December, 2017, pp. 248-252. doi:10.1145/3176653.3176711.

[156] J. Geng and J. Zhang, "A new method to solve the "boundary effect" of adaptive random testing," in *Proceedings of the International Conference on Educational and Information Technology (ICEIT)*, Chongqing, China, 2010, pp. V1-298-V1-302. doi:10.1109/ICEIT.2010.5607704.

[157] T. Y. Chen, D. H. Huang, T. H. Tse, and Z. Yang, "An innovative approach to tackling the boundary effect in adaptive random testing," in *Proceedings of the 40th Annual Hawaii International Conference on System Sciences, 2007. HICSS 2007*, Waikoloa, HI, USA, 2007, pp. 262a-262a. doi:10.1109/HICSS.2007.67.

[158] T. Y. Chen, F.-C. Kuo, and H. Liu, "Application of a failure driven test profile in random testing," *IEEE Transactions on Reliability,* vol. 58, no. 1, pp. 179-192, 2009. doi:10.1109/TR.2008.2011687.

[159] F.-C. Kuo, T. Y. Chen, H. Liu, and W. K. Chan, "Enhancing adaptive random testing for programs with high dimensional input domains or failure-unrelated parameters," *Software Quality Journal,* vol. 16, no. 3, pp. 303-327, 2008. doi:10.1007/s11219-008-9047-6.

[160] K. P. Chan, D. Towey, T. Y. Chen, F.-C. Kuo, and R. Merkel, "Using the information: incorporating positive feedback information into the testing process." Presented at the *Eleventh Annual International Workshop on Software Technology and Engineering Practice*, Amsterdam, Netherlands, September 2003, pp. 71-76. doi:10.1109/STEP.2003.38.

[161] K. P. Chan, T. Y. Chen, and D. Towey, "Normalized restricted random testing." Presented at the *Reliable Software Technologies (Ada-Europe 2003)*, Toulouse, France, June 2003, pp. 639-639. doi:10.1007/3-540-44947-7_28.

[162] H. Liu, X. Xie, J. Yang, Y. Lu, and T. Y. Chen, "Adaptive random testing by exclusion through test profile," in *Proceedings of the 10th International Conference on Quality Software (QSIC 2010)* Zhangjiajie, China, 2010, pp. 92-101. doi:10.1109/QSIC.2010.61.

[163] K. P. Chan and D. Towey, "Probabilistic adaptive random testing," in *Proceedings of the Sixth International Conference on Quality Software (QSIC'06)*, Beijing, China, 2006, pp. 274-280. doi:10.1109/QSIC.2006.48.

[164] H. Liu, X. Xie, J. Yang, Y. Lu, and T. Y. Chen, "Adaptive random testing through test profiles," *Software: Practice and Experience,* vol. 41, no. 10, pp. 1131-1154, 2011. doi:10.1002/spe.1067.

[165] H. Ackah-Arthur, J. Chen, J. Xi, M. Omari, H. Song, and R. Huang, "A Cost-Effective Adaptive Random Testing Approach by Dynamic Restriction," *IET Software,* vol. 12, no. 6, pp. 489-497, Dec. 2018. doi:10.1049/iet-sen.2017.0208.

[166] T. Y. Chen, F.-C. Kuo, and H. Liu, "Distributing test cases more evenly in adaptive random testing," *Journal of Systems and Software,* vol. 81, no. 12, pp. 2146-2162, 2008. doi:10.1016/j.jss.2008.03.062.

[167] T. Y. Chen and D. H. Huang, "Adaptive random testing by localization," in *Proceedings of the 11th Asia-Pacific Software Engineering Conference*, Busan, South Korea, 2004, pp. 292-298. doi:10.1109/APSEC.2004.17.

[168] J. Mayer, "Adaptive Random Testing by Bisection with Restriction," in *Proceedings of the International Conference on Formal Engineering Methods, 2005 (ICFEM 2005)*, Manchester, UK, November, 2005, pp. 251-263. doi:10.1007/11576280_18.

[169] C. Mao and X. Zhan, "Towards an Improvement of Bisection-Based Adaptive Random Testing." Presented at the *24th Asia-Pacific Software Engineering Conference (APSEC)* Nanjing, China, December 2017, pp. 689-694. doi:10.1109/APSEC.2017.86.

[170] K. K. Sabor and S. Thiel, "Adaptive random testing by static partitioning," in *Proceedings of the Proceedings of the 10th*

*International Workshop on Automation of Software Test*, Florence, Italy, May, 2015*,* pp. 28-32. doi:10.1109/AST.2015.13.

[171] M. Rezaalipour, L. Talebsafa, and M. Vahidi-Asl, "Arselda: An Improvement on Adaptive Random Testing by Adaptive Region Selection," in *Proceedings of the 8th International Conference on Computer and Knowledge Engineering (ICCKE)*, Mashhad, Iran, Oct. 25-26, 2018*,* pp. 73-78. doi:10.1109/ICCKE.2018.8566625.

[172] T. Y. Chen, F.-C. Kuo, R. G. Merkel, and S. P. Ng, "Mirror adaptive random testing," *Information and Software Technology,* vol. 46, no. 15, pp. 1001-1010, 2004. doi:10.1016/j.infsof.2004.07.004.

[173] R. Huang, H. Liu, X. Xie, and J. Chen, "Enhancing mirror adaptive random testing through dynamic partitioning," *Information and Software Technology,* vol. 67, pp. 13-29, 2015. doi:10.1016/j.infsof.2015.06.003.

[174] T. Y. Chen, D. H. Huang, and F.-C. Kuo, "Adaptive random testing by balancing," in *Proceedings of the 2nd international workshop on Random testing: co-located with the 22nd IEEE/ACM International Conference on Automated Software Engineering (ASE 2007)*, Atlanta, GA, USA, 2007*,* pp. 2-9. doi:10.1145/1292414.1292418.

[175] G. Klir and B. Yuan, *Fuzzy sets and fuzzy logic*: Prentice Hall, New Jersey, 1995.

[176] J. Mayer, "Lattice-based adaptive random testing," in *Proceedings of the Proceedings of the 20th IEEE/ACM International Conference on Automated software engineering*, 2005*,* pp. 333-336. doi:10.1145/1101908.1101963.

[177] T. Y. Chen, D. H. Huang, F.-C. Kuo, R. G. Merkel, and J. Mayer, "Enhanced lattice-based adaptive random testing," in *Proceedings of the Proceedings of the 2009 ACM symposium on Applied Computing*, 2009*,* pp. 422-429. doi:10.1145/1529282.1529376.

[178] T. Y. Chen and R. G. Merkel, "Efficient and effective random testing using the Voronoi diagram," in *Proceedings of the Australian Software Engineering Conference (ASWEC'06)*, Sydney, NSW, Australia, 2006, p. 6 pp. doi:10.1109/ASWEC.2006.25.

[179] Z.-W. Hui and S. Huang, "MD-ART: a test case generation method without test oracle problem," in *Proceedings of the 1st International Workshop on Specification, Comprehension, Testing, and Debugging of Concurrent Programs*, Singapore, Singapore, September, 2016*,* pp. 27-34. doi:10.1145/2975954.2975959.

[180] T. Y. Chen, J. W. Ho, H. Liu, and X. Xie, "An innovative approach for testing bioinformatics programs using metamorphic testing," *BMC bioinformatics,* vol. 10, no. 1, pp. 24, 2009. doi:10.1186/1471-2105-10-24.

[181] B. Selman and C. P. Gomes, "Hill-climbing Search," *Encyclopedia of Cognitive Science*, January 2006. doi:10.1002/0470018860.s00015.

[182] H. Hemmati, A. Arcuri, and L. Briand, "Achieving scalable model-based testing through test case diversity," *ACM Transactions on Software Engineering and Methodology (TOSEM),* vol. 22, no. 1, pp. 6, February 2013. doi:10.1145/2430536.2430540.

[183] M. Z. Iqbal, A. Arcuri, and L. Briand, *Automated System Testing of Real-Time Embedded Systems Based on Environment Models*, Simula Research Laboratory (Software V&V Laboratory), 2011.

[184] J. Bo, L. Xiang, and G. Xiaopeng, "MobileTest: A tool supporting automatic black-box test for software on smart mobile devices," in *Proceedings of the Second International Workshop on Automation of Software Test*, Minneapolis, USA, May, 2007*,* p. 8. doi:10.1109/AST.2007.9.

[185] S.-H. Shin, S.-K. Park, K.-H. Choi, and K.-H. Jung, "Normalized adaptive random test for integration tests," in *Proceedings of the 34th IEEE Annual Computer Software and Applications Conference Workshops (COMPSACW)*, 2010*,* pp. 335-340. doi:10.1109/COMPSACW.2010.65.

[186] L. Shan and H. Zhu, "Generating structurally complex test cases by data mutation: A case study of testing an automated modelling tool," *The Computer Journal,* vol. 52, no. 5, pp. 571-588, August 2009. doi:10.1093/comjnl/bxm043.

[187] M. Turner, B. Kitchenham, D. Budgen, and O. P. Brereton, "Lessons learnt undertaking a large-scale systematic literature review," in *Proceedings of the Proceedings of the 2008 International Conference on Evaluation and Assessment in Software Engineering (EASE 2008)*, Italy, June, 2008*,* pp. 110-118.